\documentclass[preprint,onecolumn,nofootinbib]{revtex4}
\usepackage[colorlinks=true,linkcolor=blue,urlcolor=blue,filecolor=black,citecolor=red,pdfstartview=FitV,pdftitle={},pdfsubject={},pdfkeywords={},pdfpagemode=None,bookmarksopen=true]{hyperref}
\usepackage{graphicx}
\usepackage{amsmath}
\usepackage{amssymb,ulem}
\usepackage{color}%
\usepackage{tikz}
\usepackage{dcolumn}
\usepackage{bbold}
\usepackage{xcolor}
\usepackage{setspace}

\begin{document}
\title{Quasi-normal modes of the Einstein-Maxwell-aether Black Hole}
\author{Wei Xiong $^{1}$}
\email{phyxw@stu2019.jnu.edu.cn}
\author{Peng Liu $^{1}$}
\email{phylp@email.jnu.edu.cn}
\author{Cheng-Yong Zhang $^{1}$}
\email{zhangcy@email.jnu.edu.cn}
\author{Chao Niu $^{1}$}
\email{niuchaophy@gmail.com}
\thanks{corresponding author}
\affiliation{
  $^1$ Department of Physics and Siyuan Laboratory, Jinan University, Guangzhou 510632, China
}

\begin{abstract}

We study the quasi-normal modes of the charged scalar perturbations in the background of the Einstein-Maxwell-aether black hole through three methods (WKB method, continued fraction method, generalized eigenvalue method). Then we propose the specific treatment for the generalized effective potential with $\omega$-dependence and the complete procedure of transforming calculation continued fraction method into finding the zero point of the corresponding complex function numerically. These methods are valid because the results from different methods are consistent. We also investigate the allowed region of the second kind aether black hole among the system parameters ($c_{13},c_{14},Q$). Finally we show the existence of quasi-resonances of massive perturbation for Einstein-Maxwell-aether black hole even with large aether parameter.

\end{abstract}
\maketitle
\tableofcontents

\section{Introduction}
\label{sec:introduction}

Quasi-normal modes are the eigenvalues of dissipative systems arising by the perturbations of additional fields or metric under the background of black hole \cite{Berti2009,Kokkotas1999,Konoplya2011,Chandrasekhar1985}. The energy dissipation of this system comes from the purely outgoing boundary condition (ingoing waves at horizon and outgoing waves at spatial infinity). The associated linear differential equation,  called perturbed equation, generates the non-Hermitian eigenvalue problem because of the non-time symmetry evolution.
In astrophysics, quasi-normal modes describe the ringdown phase of black hole mergers and allow new tests of general relativity \cite{Berti2009}. For theoretical interest, these modes were used to probe various properties of black holes with respect to quantum gravity and investigate the strongly coupled quantum field theories due to the gauge-gravity duality \cite{Berti2009,Eniceicu2020}. Currently there are gravitational wave signals detected by LIGO \cite{Abbott2016a,Abbott2016b}. However, various modified gravity theories are not excluded by the current observed data in the gravitational and electromagnetic spectra because of the large uncertainty of the determination of the mass and angular momentum of black holes \cite{Churilova2020}.
    
The modified gravity theory we are interested in is the Einstein-Maxwell-aether theory, which belongs to the Lorentz violation (LV) models. The reason for introducing Lorentz violation is that Lorentz invariance may not be an exact symmetry at all energies \cite{Mattingly2005}. Different quantum gravity theories have investigated the possibility of the existence of Lorentz violation and the non-commutative field theory, one of the high energy models of spacetime structure, even contains Lorentz violation explicitly \cite{Douglas2001}. In Einstein-Maxwell-aether theory, the Lorentz violating terms are added to the gravity sector in the dynamical framework \cite{Berti2009}. This model assumes that each point of spacetime introduces a preferred timelike direction, marked by an aether vector field $u^{a}$. This assumption caused the Lorentz symmetry to be broken down to a rotation subgroup. Many interesting phenomena have been found in this theory \cite{Zhu2019,Ding2015a,Lin2016,Jacobson2001,Ding2015,Ghosh2021}, e.g., a superluminal group velocity is allowed for the modified scalar field \cite{Jacobson2001}. The corresponding light-cones therefore can be completely flat and the causality is not violated by the superluminal phenomena. On the other hand, there are black hole solutions in aether theory \cite{Ding2015}. A three-dimensional spacelike hypersurface, called universal horizon, replaces the killing horizon as the event horizon because of the existence of superluminal particle, i.e., because only the universal horizon can trap arbitrarily fast excitations \cite{Bhatta2014}. Furthermore, the Einstein-Maxwell-aether theory here introduces the extra source-Free Maxwell field than pure aether theory. For black holes, the interaction between photon and deformed aether may induce new dynamo-optical effects \cite{Ding2015}. The quasi-normal modes of the uncharged aether theory have been investigated by Konoplya \cite{Konoplya2007}, Ding \cite{Ding2019,Ding2017} and Churilova \cite{Churilova2020} for scalar or gravitational perturbations. Churilova \cite{Churilova2020} argued that the perturbation of the energy-momentum tensor of the aether field should be taken into account for the calculation of the Einstein-aether gravitational perturbations.

In this paper, we first calculate the quasi-normal frequencies of the charged massless scalar perturbation in the background of the Einstein-Maxwell-aether black hole. For this charged background, the charged scalar field is naturally introduced to couple with the electromagnetic field \cite{Eniceicu2020,Konoplya2013}, which can better reveal the new effect of the electromagnetic field on the background of the aether black hole. This perturbation field is necessary for studying the perturbations of charged particles in scalar electrodynamics in the curved charged background \cite{Konoplya2011}, which is the Einstein-Maxwell-aether black hole in our cases.

We also confirm the existence of the quasi-resonances under the charged aether black hole background, by calculating the massive charged perturbations under these backgrounds. 
The quasi-resonance is the arbitrary long-living mode with real frequency, which is due to the nonzero value of at least one of the boundaries of the effective
potential \cite{Konoplya2005,Konoplya2006a,Konoplya2006b,Konoplya2013,Wu2015,Zinhailo2018,Abdalla2019,Konoplya2019a,Zinhailo2019,Konoplya2018,Churilova2020a,Churilova2020,Churilova2019}. 

The paper is organized as follows. In section \ref{section2}, we introduce the Einstein-Maxwell-aether theory and two black hole solutions. In section \ref{section3}, we specify three different methods for the calculations. The results of the quasi-normal frequencies are presented in section \ref{section4} and the discussion of quasi-resonances is shown in section \ref{section5}. We conclude in section \ref{section6}.

\section{Einstein-Maxwell-aether Black Hole}\label{section2}
In this section we briefly review the Einstein-Maxwell-aether theory and investigate the scalar perturbation around black hole solutions.

\subsection{Einstein-Maxwell-aether theory}
The action of Einstein-Maxwell-aether theory is \cite{Ding2015}
\begin{equation}
  S=\int d^{4}x \sqrt{-g} \left[ \frac{1}{16 \pi G_{\textrm{\ae}}} 
  (\mathcal{R}+\mathcal{L}_{\textrm{\ae}}) + \mathcal{L}_{M} \right],
\end{equation}
where the $g$ is the determinant of the metric $g_{\mu\nu}$ and $R$ the Ricci scalar. The constant $G_{\textrm{\ae}}$ denotes the Newton's gravitational constants $G_{N}$ by $G_{\textrm{\ae}}=(1-c_{14}/2)G_{N}$, which is obtained by the renormalization of the total energy of Einstein-aether theory \cite{Eling2006}. The aether Lagrangian is
\begin{eqnarray}
  \mathcal{L}_{\textrm{\ae}} & = & -Z^{ab}{}_{cd} (\nabla_{a}u^{c}) 
  (\nabla_{b}u^{d}) + \lambda (u^{2}+1),\nonumber\\
  Z^{ab}{}_{cd} & = & c_{1} g^{ab} g_{cd} + c_{2} \delta^{a}{}_{c} 
  \delta^{b}{}_{d} + c_{3} \delta^{a}{}_{d} \delta^{b}{}_{d}-
  c_{4} u^{a} u^{b} g_{cd},
\end{eqnarray}
where $c_{i} (i=1,2,3,4)$ are coupling constants of the theory and the $\lambda(u^{2}+1)$ term constraints the vector field to satisfy the normalization condition $u^{2}=-1$. The source-free Maxwell Lagrangian is 
\begin{eqnarray}
  \mathcal{L}_{M} &=& - \frac{1}{16\pi G_{\textrm{\ae}}} \mathcal{F}_{ab} \mathcal{F}^{ab}, \nonumber\\
  \mathcal{F}_{ab}&=& \nabla_{a}\mathcal{A}_{b}-\nabla_{b}\mathcal{A}_{a}.
\end{eqnarray}

The observational and theoretical bounds of the coupling constants $c_{i}$ have been investigated by requiring the absence of gravitational Cherenkov radiation for theoretical constants \cite{Yagi2014,Jacobson2001,Jacobson2007,Jacobson2008}. Because of the theoretical interest of this LV gravity theory, we impose the following constraints \cite{Berglund2012},
\begin{equation}
  0\leq c_{14} <2, \quad 2+c_{13}+3c_{2}>0, \quad 0\leq c_{13}<1,
  \label{eq:theoretical constraints}
\end{equation}
where $c_{ij} \equiv c_{i}+c_{j}$.    

\subsection{Black hole solutions}
In asymptotical flat spherically symmetric spacetime, the static metric for Einstein-Maxwell-aether black hole is given by
\begin{equation}
  ds^{2}=-f(r)dt^{2}+\frac{dr^{2}}{f(r)}+r^{2}(d\theta^{2}+\sin^{2}\theta d\varphi^{2}).\label{eq:metric}
\end{equation}

There are two kinds of exact solutions for the cases $c_{14}=0,\;c_{123}\neq 0$ and $c_{123}=0,\;c_{14}\neq 0$ \cite{Ding2015}. They represent two different behaviors of propagation speed of spin-0 mode with respect to linearized aether-metric perturbations around flat spacetime respectively \cite{Jacobson2004,Bhatta2014}. In the first kind aether black hole ($c_{14}=0,\;c_{123}\neq 0$), the metric function is
\begin{eqnarray}
  f(r)&=&1-\frac{r_{0}}{r}+\frac{Q^{2}}{r^2}+\frac{c_{13}\; B}{(1-c_{13})r^{4}}, \nonumber\\
  B&=&\frac{(r_{0}-\sqrt{-32Q^{2}+9r_{0}^2})(3r_{0}+\sqrt{-32Q^{2}+9r_{0}^2})^{3}}{4096},
\end{eqnarray}
where ADM mass is given by $r_{0}/2G_{\textrm{\ae}}$, with the constraint of charged $Q\leq r_{0}/2$ which is obtained by requiring regularity of the aether theory for each point in the spacetime \cite{Ding2015}. This constraint is the same as that given in the Reissner-Nordström black hole. The metric function shows that the aether correction term is added as $\mathcal{O}(1/r^{4})$ in the metric and this correction term vanishes when $c_{13}\rightarrow 0$.

The metric function of the second kind aether black hole ($c_{123}=0,\;c_{14}\neq 0$)is given by
\begin{eqnarray}
  f(r)&=&1-\frac{r_{0}}{r}-\frac{r_{u}(r_{0}+r_{u})}{r^{2}}, \nonumber\\
  r_{u}&=&\frac{r_{0}}{2} \left( \sqrt{\frac{2-c_{14}}{2(1-c_{13})}-\frac{4Q^{2}}{(1-c_{13})r_{0}^{2}}}-1\right),
  \label{eq:metric function for second kind}
\end{eqnarray}
with the following constraints
\begin{eqnarray}
  Q &\leq& \frac{r_{0}}{2} \sqrt{c_{13}-\frac{c_{14}}{2}} , \nonumber\\
  c_{13}&\geq&\frac{c_{14}}{2}.\label{eq:constraint for c123=0}
\end{eqnarray}

\subsection{Massless charged scalar perturbation}
Here, we present a massless charged scalar perturbation around black hole solutions (\ref{eq:metric}). For massless charged scalar perturbation, the first-order perturbation $\mathcal{O}(\epsilon)$ of the scalar field in the given background perturbs the spacetime at the second order $\mathcal{O}(\epsilon^{2})$ \cite{Brito2015}, i.e., to leading order, we can treat this scalar perturbation as a probe into the background with a fixed geometry. Therefore the following perturbation equation is
\begin{equation}
  D_{\mu}D^{\mu}\Phi=0,
  \label{eq:charged scalar}
\end{equation}
where $D_{\mu}\equiv \nabla_{\mu}-ieA_{\mu}$, $A_{\mu}$ is the electromagnetic potential four-vector above, and $e$ the test charge of the scalar field.

The source-free Maxwell field $\mathcal{F}_{ab}$ has been imposed in \cite{Ding2015} and the vector-potential is given by
\begin{equation}
  A=-\frac{Q}{r}dt.
\end{equation}

The equation of motion (\ref{eq:charged scalar}) can be separated into
\begin{equation}
  \Phi(t,r,\theta,\varphi)=\sum_{lm} \int d\omega \; e^{-i\omega t} \frac{\phi(r)}{r} Y_{lm}(\theta,\varphi),
\end{equation}
where $Y_{lm}(\theta,\varphi)$ is the spherical harmonics for the 2-sphere $S^{2}$. Hence (\ref{eq:charged scalar}) remains the radial equation
\begin{equation}
  \phi(r)\left[(eQ-r\omega)^{2} -r f(r) \left( l(l+1)r + f'(r) \right) \right] + r^{2}f(r)f'(r) \phi '(r)+r^{2}f^{2}(r)\phi ''(r)=0.
  \label{eq:separate}
\end{equation}
The main difference between this equation and the uncharged case is the introduction of $(eQ-r\omega)^{2}$ term, which brings the first power term of $\omega$ and modifies the boundary behavior at the horizon.

\section{The Methods}\label{section3}
In this section we summarize the three methods we will use in our computation of quasi-normal modes.
\subsection{The WKB method}
\begin{figure}[htbp]
  \centering
  \includegraphics[width = 0.35\textwidth]{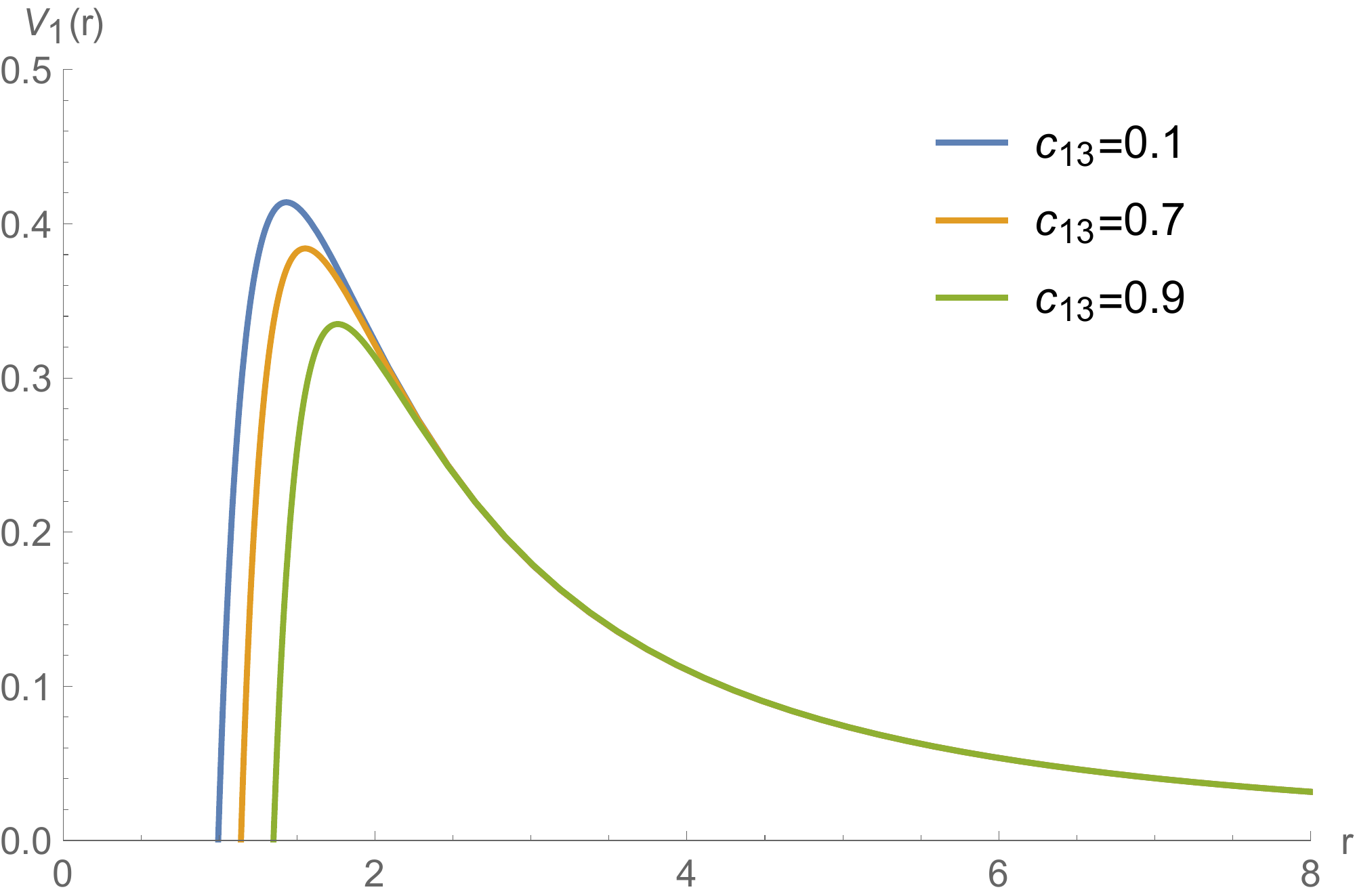}
  \includegraphics[width = 0.35\textwidth]{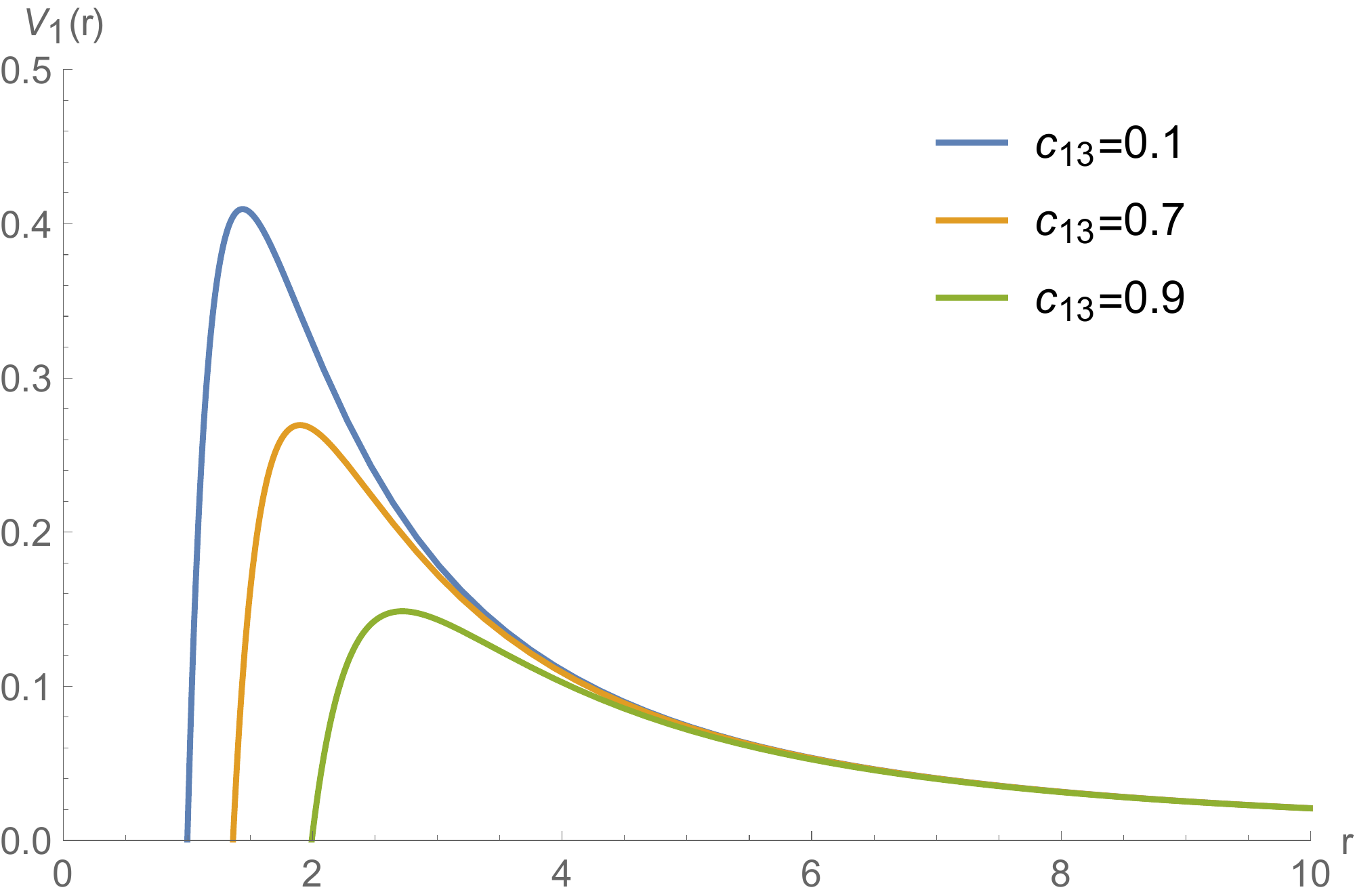}
  \caption{The effective potential for the first kind aether black hole (left) and the second kind aether black hole (right) with $l=1,\,e=0.1,\,Q=0.1,\,r_{0}=1$. }
  \label{fig:effective potential}
\end{figure}

For WKB method, the equation (\ref{eq:separate}) is conventionally reduced to the Schrödinger-like equation by taking the tortoise coordinate $dr_{*}=dr/f(r)$,
\begin{eqnarray}
  \frac{d^{2}\phi(r)}{dr_{*}^{2}}&=&\left[ V(\omega,r)-\omega^{2} \right] \phi(r), \label{eq:Schrodinger-like}\\
  V(\omega,r)&=&\frac{2eQ\omega}{r} + \frac{e^{2}Q^{2}}{r^{2}} -l(l+1)f(r)-\frac{f(r)f'(r)}{r} \label{eq:effective potential},
\end{eqnarray}
where $r_{*}$ ranges from $-\infty$ at the horizon to $+\infty$ at the radial infinity. This Schrödinger-like equation has an effective potential barrier (Fig. \ref{fig:effective potential}) between the horizon and the radial infinity for both first and second kind aether black hole, where the $\omega$-dependence is assumed to be $\omega =1$. These effective potentials depend on the frequency $\omega$ for charged scalar perturbation, which is different from the uncharged case \cite{Iyer1987a,Iyer1987b,Ding2019}. However the WKB method is also suitable for this effective potential \cite{Iyer1987a,Konoplya2002,Konoplya2019} when the quasi-normal modes satisfy the condition $\textrm{Re}(\omega)>0$ \cite{Konoplya2019}. The scheme to solve this problem will be discussed in detail below. 

The WKB method can semi-analytically calculate the quasi-normal modes which satisfy the suitable boundary conditions (ingoing for horizon and outgoing for spatial infinity). The first-order WKB was first used by Schutz and Will \cite{Schutz1985}, and the third-order WKB was given by Iyer and Will \cite{Iyer1987a} two years later. After that this method was extended to the sixth-order by Konoplya \cite{Konoplya2002} and to the thirteenth-order by Matyjasek and Opala \cite{Matyjasek2017}. The higher-order WKB formula is given by \cite{Konoplya2019}
\begin{eqnarray}
  0 &=& U_{0}(\omega)+A_{2}(\mathcal{K}^{2})+A_{4}(\mathcal{K}^{2})+A_{6}(\mathcal{K}^{2})+\ldots \nonumber \\
  &-& i \mathcal{K} \sqrt{-2U_{2}(\omega)} (1+A_{3}(\mathcal{K}^{2})+A_{5}(\mathcal{K}^{2})+A_{7}(\mathcal{K}^{2})+\ldots), 
  \label{eq:WKB1}
\end{eqnarray}
where
\begin{equation}
  \mathcal{K}=n+\frac{1}{2},\, U_{0}(\omega)=U(\omega,r_{\textrm{max}}),\, \left.  U_{2}(\omega)=\frac{d^{2}U(\omega,r)}{dr_{*}^{2}} \right|_{r=r_{\textrm{max}}},\,  \left.  U_{3}(\omega)=\frac{d^{3}U(\omega,r)}{dr_{*}^{3}} \right|_{r=r_{\textrm{max}}},\, \ldots.
\end{equation}
The $r_{\textrm{max}}$ denotes the value of coordinate $r$ where the effective potential (\ref{eq:effective potential}) reaches its maximum and the $n$ is the overtone number. The $A_{i}(\mathcal{K}^{2})$ is the correction of order $i$ which depends on $\mathcal{K}^{2}$ and the values $U_{2},U_{3},\ldots$ of higher derivatives of $U_{0}$. For charged scalar perturbation, the $U(\omega,r)$ can be expended by $V(\omega,r)-\omega^{2}$. Hence the WKB formula becomes
\begin{eqnarray}
  \omega^{2} &=& V_{0}(\omega)+A_{2}(\mathcal{K}^{2})+A_{4}(\mathcal{K}^{2})+A_{6}(\mathcal{K}^{2})+\ldots \nonumber \\
  &-& i \mathcal{K} \sqrt{-2V_{2}(\omega)} (1+A_{3}(\mathcal{K}^{2})+A_{5}(\mathcal{K}^{2})+A_{7}(\mathcal{K}^{2})+\ldots)\, , 
  \label{eq:WKB2}
\end{eqnarray}
where the meanings of $V_{0},V_{2},\ldots$ are similar to $U_{0},U_{2},\ldots$ above. 

The accuracy of third-order WKB has been verified for the low-lying modes $n<l$ \cite{Iyer1987b,Kokkotas1988} and has significant error for $n>l$. However the result can not be improved by simply increasing the WKB formula order due to the asymptotical convergence of the WKB method \cite{Hatsuda2019}. The Pad\'{e} approximants can be used to improve the accuracy of the higher-order WKB method \cite{Konoplya2019}. This approach starts by defining a polynomial $P_{k}(\epsilon)$
\begin{eqnarray}
  P_{k}(\epsilon) &=& V_{0}(\omega)+A_{2}(\mathcal{K}^{2})\epsilon^{2}+A_{4}(\mathcal{K}^{2})\epsilon^{4}+A_{6}(\mathcal{K}^{2})\epsilon^{6}+\ldots \nonumber \\
  &-& i \mathcal{K} \sqrt{-2V_{2}(\omega)} (\epsilon+A_{3}(\mathcal{K}^{2})\epsilon^{3}+A_{5}(\mathcal{K}^{2})\epsilon^{5}+\ldots)\, , 
  \label{eq:pade}
\end{eqnarray}
and it returns to WKB formula (\ref{eq:WKB2}) by taking $\epsilon=1$,
\begin{equation}
  \omega^{2}=P_{k}(1).
\end{equation}

Next we can construct a family of rational functions called Pad\'{e} approximants
\begin{equation}
  P_{\tilde{n}/\tilde{m}}(\epsilon)=\frac{Q_{0}+Q_{1}\epsilon+\ldots+Q_{\tilde{n}}\epsilon^{\tilde{n}}}{R_{0}+R_{1}\epsilon+\ldots+R_{\tilde{m}}\epsilon^{\tilde{m}}}
  \label{eq:pade approximants}
\end{equation}
with $\tilde{n}+\tilde{m}=k$. The divergence between Pad\'{e} approximants and pure WKB formula is given by $O(\epsilon^{k+1})$ near $\epsilon=0$.

However the turning point $r_{\textrm{max}}$ can not be evaluated directly due to the $\omega$-dependence of the effective potential $V(\omega,r)$ and the correction $A_{i}(\mathcal{K}^{2})$ will become a rather complicated function of $\omega$ with  increasing of $i$. In more detail, Matyjasek and Opala \cite{Matyjasek2017} showed the number of terms in $A_{i}$, where the $A_{6}$ has 294 terms and  the $A_{13}$ has even 22050 terms. Moreover, the Pad\'{e} approximants further increase the computational complexity.

If we treat $\omega$ as real-value, the turning point $r_{\textrm{max}}$ of effective potential becomes a numerical function of $\omega$ when fixes all the other parameters \cite{Konoplya2002a}. The following procedure is given by Konoplya \cite{Konoplya2002}: substituting the numerical function into the equation (\ref{eq:WKB2}) and the quasi-normal frequency is obtained by finding the root of this equation. More specifically, we move $\omega^{2}$ to the right of the equation (\ref{eq:WKB2}) and regard the right part as a numerical complex function on the complex $\omega$ plane. For a given $\omega$, the $r_{\textrm{max}}$ can be found and then the value of the numerical complex function is obtained through the open $Mathematica$ package \cite{package1}. Through an iterative program, the quasi-normal frequency is determined by the approximation of the function value to zero.

In this paper, we use sixth-order WKB formula with pad\'{e} approximants $P_{5/1}(1)$ to reproduce the results in \cite{Churilova2020} and calculate the charged case.

\subsection{The continued fraction method}
Since Leaver first reported the results of quasi-normal modes calculated by the continued fraction method \cite{Leaver}, this method has been widely used in the study of scalar perturbations under different black hole theories. However, it is difficult to obtain a complete three-term recurrence formula including all theoretical parameters. We use the numerical program to obtain the three-term recurrence formula, which will be expanded in detail below.

Equation (\ref{eq:separate}) can be written in a different form
\begin{equation}
  f(r) \left( p(r) \frac{d^{2}}{dr^{2}} +  \frac{d}{dr} + q(r,\omega) \right) \phi(r)=0.
  \label{eq:equation of continued fraction method}
\end{equation}
When we fix all the parameters ($e,\:Q,\:c_{13},\:c_{14},\:r_{0},\:l$), the Frobenius series can be constructed as \cite{Konoplya2011}
\begin{equation}
  e^{-I(\omega)\: r} r^{-I(\omega)} \left( \frac{r-r_{h}}{r} \right)^{H(\omega)} \sum^{\infty}_{k=0} b_{k} \left( \frac{r-r_{h}}{r} \right)^{k},
  \label{eq:Frobenius series}
\end{equation}
where $I(\omega),H(\omega)$ are functions of purely $\omega$ and defined by the boundary conditions of Eq. (\ref{eq:equation of continued fraction method}). For instance, both $I(\omega) \textrm{ and } H(\omega)$ are $-i\omega$ for massless scalar perturbation of Schwarzschild black hole.

The Frobenius Series (\ref{eq:Frobenius series}) can be truncated to N and substituted into (\ref{eq:equation of continued fraction method}). The N-term recurrence relation for the coefficients $b_{k}$ is
\begin{equation}
  \sum^{\textrm{min}(N-1,i)}_{j=0} c_{j,i}^{(N)} b_{i-j} =0, \qquad \textrm{for} \; i>0.
  \label{eq:N term recurrence relation}
\end{equation}

Then, the Gaussian eliminations allows one to reduce the N-term recurrence relation to the three-term recurrence relation \cite{Konoplya2011}
\begin{eqnarray}
  c^{3}_{0,i}b_{i}+c_{1,i}^{(3)}b_{i-1}+c_{2,i}^{(3)}b_{i-2}&=&0, \quad \textrm{for} \; i>1, \nonumber \\
  c^{(3)}_{0,1}b_{1}+c_{1,1}^{(3)}b_{0} &=& 0.
  \label{eq:3 term recurrence relation}
\end{eqnarray}

In our cases, the derivations of the three-term recurrence relations can be executed from Eq. (\ref{eq:N term recurrence relation}) by the $Mathematica$ program step by step. Here, we can find $b_{1}/b_{0}$ from the recurrence relation (\ref{eq:3 term recurrence relation}) in two ways:
\begin{equation}
  \frac{b_{1}}{b_{0}}= -\frac{c_{1,1}^{(3)}}{c^{(3)}_{0,1}}=-\frac{c^{(3)}_{2,2}}{c^{(3)}_{1,2}- \frac{c^{(3)}_{0,2} c^{(3)}_{2,3}}{c^{(3)}_{1,3}-
  \frac{c^{(3)}_{0,3}c^{(3)}_{2,4}}{c^{(3)}_{1,4}-\cdots } } } .
\end{equation}
And the final equation with respect to the coefficients of the three-term recurrence relations is given by
\begin{equation}
  c^{(3)}_{1,1}-\frac{c^{(3)}_{0,1} c^{(3)}_{2,2}}{c^{(3)}_{1,2}- \frac{c^{(3)}_{0,2}c^{(3)}_{2,3}}{c^{(3)}_{1,3}- \cdots }}=0.
  \label{eq:final equation of continued fraction}
\end{equation}

Recall that Eq. (\ref{eq:final equation of continued fraction}) holds only if $\omega$ is the quasi-normal frequency. We choose this equation as the basis for judging whether $\omega$ is the quasi-normal frequency. The procedure is the following: determining the value of $\omega$ and substituting it into Eq. (\ref{eq:Frobenius series}); then the coefficients $c_{j,i}^{(N)}$ of Eq. (\ref{eq:N term recurrence relation}) become complex numbers, which can be conveniently reduced to the three-term recurrence relation (\ref{eq:3 term recurrence relation}) by the program; the coefficient of the three-term recurrence relation can be used to construct the left part of Eq. (\ref{eq:final equation of continued fraction}), that is, finally we obtain the result as a complex number, which represents the left part of Eq. (\ref{eq:final equation of continued fraction}). If this complex number is zero, we can conclude that the $\omega$ is the quasi-normal frequency. Therefore we turn the problem into searching the zero point of the complex function (the left part of Eq. (\ref{eq:final equation of continued fraction})) numerically on the complex $\omega$ plane.

\subsection{The generalized eigenvalue method}
This numerical method developed by Jansen \cite{Jansen2017} is finding the quasi-normal modes by discretizing the perturbation equation and solving the resulting generalized eigenvalue equation. This method is convenient to work under the ingoing Eddington-Finkelstein coordinate
\begin{equation}
  ds^{2}=-f(z)dv^{2}-2 z^{-2} dv dz+ z^{-2} d\Omega^{2}_{2},
  \label{eq:ingoing Eddington metric}
\end{equation}
where $z \equiv 1/r$ and $v \equiv t+r_{*} $. Substituting it into (\ref{eq:charged scalar}) and separating the result equation by
\begin{equation}
  \Phi(v,z,\theta,\varphi)=\sum_{lm} \int d\omega \; e^{-i\omega v} \phi(z) Y_{lm}(\theta,\varphi).
\end{equation}
The remain radial equation is
\begin{equation}
  (lz+l^{2}z-ieQz+2i\omega) \phi(z) + \left(2iz(eQz-\omega)-z^{3}f'(z)\right) \phi'(z)-z^{3}f(z)\phi''(z)=0.
  \label{eq:charged scalar for Eddington}
\end{equation}

Then we need to apply appropriate boundary behaviors to satisfy the ingoing boundary condition near the horizon and the outgoing boundary condition near the infinity. The main idea of this operation is rescaling the equation to make the inappropriate solution (the non-normalizable solution or the solution which does not satisfy the boundary conditions) pathological, diverging, and rapidly oscillating. According to this idea, We redefine
\begin{equation}
  \phi(z) \rightarrow e^{2\omega i/z}\: z^{-2\omega i} \phi(z),
\end{equation}
and the final equation becomes
\begin{eqnarray}
  \left(lz+l^{2}z-\frac{2\omega(-iz+2\omega+2z\omega)}{z}+eQ(-iz+4\omega+4z\omega) \right. & & \nonumber \\
  +\left. \frac{2\omega(2\omega+z^{2}(-i+2\omega)+z(-2i+4\omega))f(z)}{z}+2i  z(1+z) \omega f'(z) \right) \phi(z) & & \nonumber \\
  +\left(2iz(eQz-\omega)+4iz(1+z)\omega f(z)-z^{3}f'(z) \right)\phi'(z)-z^{3}f(z)\phi''(z) &=&0.
  \label{eq:improve equation}
\end{eqnarray}
The asymptotic behaviors of this equation near the boundary can be tested by plugging in the ansatz $\phi(z)=(1/r_{h}-z)^{p}$ at the horizon and $e^{p/z}z^{1-p}$ at the spatial infinity. There are only ingoing waves to the horizon and outgoing waves to the infinity, while the ingoing modes from infinity diverge. 

Next we choose the Chebyshev grid to discretize this equation and the $n$-order derivative is replacing by the $N \times N$ matrix $D_{ij}^{(n)}$ \cite{Jansen2017}. The result matrix equation of (\ref{eq:improve equation}) depends on the square of frequency $\omega$
\begin{equation}
  (\tilde{M}_{0}+\omega \tilde{M}_{1}+\omega^{2}\tilde{M}_{2})\phi=0.
  \label{eq:square omega}
\end{equation}
The generalized eigenvalue equation only requires first power of frequency, so we can define
\begin{equation}
  M_{0}=\left(
    \begin{array}{cc}
      \tilde{M}_{0} & \tilde{M}_{1}\\
      0 & \mathbb{1}
    \end{array}
        \right) \quad ,\quad 
  M_{1}= \left(
    \begin{array}{cc}
      0 & \tilde{M}_{2}\\
      -\mathbb{1} & 0
    \end{array}
        \right),
\end{equation}
where $\mathbb{1}$ is the $N$-dimensional identity matrix. These $2N \times 2N$ matrices above act on the vector $\Phi \equiv (\phi \, , \, \omega \phi)$ and the resulting equation is 
\begin{equation}
  (M_{0}+\omega M_{1})\; \Phi =0.
  \label{eq:generalized eigenvalue}
\end{equation}
The more details are presented by Jansen in \cite{Jansen2017}.

\section{Quasi-normal Modes}\label{section4}
In this section, we investigated the fundamental quasi-normal mode ($n=0$) for the charged scalar perturbation in the background of the Einstein-Maxwell-aether black hole. We focus on the lower multipole numbers ($l=0,1,2$) and set $r_{0} = 1$ by convention, which implies that the dimensionaless frequency should be taken as $\omega \rightarrow 2\omega$ \cite{Iyer1987b}.

For the results, we define a relative effect between the results of aether cases and non-aether cases (Schwarzschild black hole for uncharged scalar perturbation and RN black hole for charged scalar perturbation) of continued fraction method as
\begin{eqnarray}
  \delta_{\textrm{Re}} &=& \frac{|\textrm{Re}\; \omega_{i}-\textrm{Re}\; \omega_{0}|}{\textrm{Re}\; \omega_{0}} \times 100 \% \\
  \delta_{\textrm{Im}} &=& \frac{|\textrm{Im}\; \omega_{i}-\textrm{Im}\; \omega_{0}|}{\textrm{Im}\; \omega_{0}} \times 100 \% ,
\end{eqnarray}
where $\omega_{i}$ is the result of different $c_{13}$ and $\omega_{0}$ is the mode without aether field. 

\subsection{Spectrum of uncharged black hole}

\begin{center}
  \begin{table*}
  \begin{tabular}{p{2cm}p{4.5cm}p{1cm}p{1cm}p{4.5cm}p{1cm}p{1cm}}
  \multicolumn{7}{l}{TABLE I. Fundamental modes of the uncharged cases for the first kind aether black hole}\\
  \multicolumn{7}{l}{ with $Q=0 $, $e=0$, obtained by WKB (first line), continued fraction method (second line)}\\
  \multicolumn{7}{l}{ and generalized eigenvalue method  (third line).}\\
  \hline \hline
  \begin{tabular}{l} \\ Parameter \\ \hline $c_{13}$ \end{tabular} & 
  \multicolumn{3}{c}{
      \begin{tabular}{p{4.5cm}p{1cm}p{1cm}} 
          \multicolumn{3}{c}{$\ell=0$} \\ \hline
          $\;\;\;\;\;\;\;\;\;\;$ QNM & \multicolumn{2}{c}{Effect \% $\;\;\;\;$}\\ \hline
          $\;\;\;\;\;\;\;\;\;\;\;\;\;$ $\omega$ & $\delta_{\mathrm{Re}}$  & $\delta_{\mathrm{Im}}$
      \end{tabular}
  } &
  \multicolumn{3}{c}{
      \begin{tabular}{p{4.5cm}p{1cm}p{1cm}} 
          \multicolumn{3}{c}{$\ell=1$} \\ \hline
          $\;\;\;\;\;\;\;\;\;\;$ QNM & \multicolumn{2}{c}{Effect \% $\;\;\;\;$}\\ \hline
          $\;\;\;\;\;\;\;\;\;\;\;\;\;$ $\omega$ & $\delta_{\mathrm{Re}}$  & $\delta_{\mathrm{Im}}$
      \end{tabular}
  } \\[3pt] \hline \\[-12pt]
  \begin{tabular}{l} $0$ \\[-3pt] $ $ \\[-3pt] $ $ \end{tabular} & 
  \begin{tabular}{l} $0.110678 - 0.104424i$ \\[-3pt] $0.110455-0.104896i$ \\[-3pt] $0.110455 - 0.104896i$ \end{tabular} & 
  \begin{tabular}{l} $ $ \\[-3pt] $0$ \\[-3pt] $ $ \end{tabular} & \begin{tabular}{l} $ $ \\[-3pt] $0$ \\[-3pt] $ $ \end{tabular} & 
  \begin{tabular}{l} $0.292932-0.097660i $ \\[-3pt] $0.292936-0.097660i $ \\[-3pt] $0.292936-0.097660i $ \end{tabular} & 
  \begin{tabular}{l} $ $ \\[-3pt] $0$ \\[-3pt] $ $ \end{tabular} & \begin{tabular}{l} $ $ \\[-3pt] $0$ \\[-3pt] $ $ \end{tabular} \\[-1pt]
  \begin{tabular}{l} $0.15$ \\[-3pt] $ $  \\[-3pt] $ $ \end{tabular} & 
  \begin{tabular}{l} $0.109637-0.105590i $ \\[-3pt] $0.109300-0.106164i $ \\[-3pt] $0.109300-0.106164i $ \end{tabular} & 
  \begin{tabular}{l} $ $\\[-3pt]  $0.99 $\\[-3pt] $ $\end{tabular} & \begin{tabular}{l} $ $ \\[-3pt] $1.2 $\\[-3pt] $ $ \end{tabular} & 
  \begin{tabular}{l} $0.291163-0.098740i $\\[-3pt] $0.291167-0.098754i $ \\[-3pt] $0.291167-0.098754i $ \end{tabular} & 
  \begin{tabular}{l} $ $ \\[-3pt] $0.6$\\[-3pt] $ $ \end{tabular} & \begin{tabular}{l} $ $ \\[-3pt] $1.1 $ \\[-3pt] $ $\end{tabular} \\[-1pt]
  \begin{tabular}{l} $0.3$ \\[-3pt] $ $   \\[-3pt] $ $  \end{tabular} & 
  \begin{tabular}{l} $0.107641-0.105651i $  \\[-3pt] $0.107708-0.107620i $ \\[-3pt] $0.107708-0.107620i $  \end{tabular} & 
  \begin{tabular}{l} $ $ \\[-3pt] $2.5 $  \\[-3pt] $ $ \end{tabular} & \begin{tabular}{l} $ $ \\[-3pt] $2.6  $  \\[-3pt] $ $ \end{tabular} & 
  \begin{tabular}{l} $0.288751-0.100052i $  \\[-3pt] $0.288760-0.100062i $ \\[-3pt] $0.288760-0.100062i $ \end{tabular} & 
  \begin{tabular}{l} $ $ \\[-3pt] $1.4 $  \\[-3pt] $ $ \end{tabular} & \begin{tabular}{l} $ $ \\[-3pt] $2.5 $ \\[-3pt] $ $  \end{tabular} \\[-1pt]
  \begin{tabular}{l} $0.45$ \\[-3pt] $ $  \\[-3pt] $ $ \end{tabular} & 
  \begin{tabular}{l} $0.104550-0.107923i $  \\[-3pt] $0.105418-0.109303i $ \\[-3pt] $0.105418-0.109302i $ \end{tabular} & 
  \begin{tabular}{l} $ $ \\[-3pt] $4.5 $  \\[-3pt] $ $ \end{tabular} & \begin{tabular}{l} $ $ \\[-3pt] $4.2 $  \\[-3pt] $ $ \end{tabular} & 
  \begin{tabular}{l} $0.285297-0.101644i $ \\[-3pt] $0.285311-0.101651i $ \\[-3pt] $0.285311-0.101651i $ \end{tabular} & 
  \begin{tabular}{l} $ $ \\[-3pt] $2.6 $  \\[-3pt] $ $ \end{tabular} & \begin{tabular}{l} $ $ \\[-3pt] $4.1 $  \\[-3pt] $ $ \end{tabular} \\[-1pt]
  \begin{tabular}{l} $0.6$ \\[-3pt] $ $  \\[-3pt] $ $ \end{tabular} & 
  \begin{tabular}{l} $0.101186-0.110012i $  \\[-3pt] $0.101901-0.111241i $ \\[-3pt] $0.101901-0.111240i $ \end{tabular} & 
  \begin{tabular}{l} $ $ \\[-3pt] $7.8 $  \\[-3pt] $ $ \end{tabular} & \begin{tabular}{l} $ $ \\[-3pt] $6.0 $  \\[-3pt] $ $ \end{tabular} & 
  \begin{tabular}{l} $0.279946-0.103602i $  \\[-3pt] $0.279968-0.103611i $ \\[-3pt] $0.279968-0.103611i $ \end{tabular} & 
  \begin{tabular}{l} $ $ \\[-3pt] $4.4 $  \\[-3pt] $ $ \end{tabular} & \begin{tabular}{l} $ $ \\[-3pt] $6.1 $  \\[-3pt] $ $ \end{tabular} \\[-1pt]
  \begin{tabular}{l} $0.75$ \\[-3pt] $ $  \\[-3pt] $ $ \end{tabular} & 
  \begin{tabular}{l} $0.095375-0.112333i $  \\[-3pt] $0.095816-0.113345i $ \\[-3pt] $0.095816-0.113345i $ \end{tabular} & 
  \begin{tabular}{l} $ $ \\[-3pt] $13 $  \\[-3pt] $ $ \end{tabular} & \begin{tabular}{l} $ $ \\[-3pt] $8.1 $  \\[-3pt] $ $ \end{tabular} & 
  \begin{tabular}{l} $0.270482-0.105977i $  \\[-3pt] $0.270476-0.106006i $ \\[-3pt] $0.270476-0.106006i $ \end{tabular} & 
  \begin{tabular}{l} $ $ \\[-3pt] $7.7 $  \\[-3pt] $ $ \end{tabular} & \begin{tabular}{l} $ $ \\[-3pt] $8.5 $  \\[-3pt] $ $ \end{tabular} \\[-1pt]
  \begin{tabular}{l} $0.9$ \\[-3pt] $ $  \\[-3pt] $ $ \end{tabular} & 
  \begin{tabular}{l} $0.082006-0.114565i $ \\[-3pt] $0.081840-0.114256i $  \\[-3pt] $0.081835-0.114258i $ \end{tabular} & 
  \begin{tabular}{l} $ $ \\[-3pt] $26 $ \\[-3pt] $ $  \end{tabular} & \begin{tabular}{l} $ $ \\[-3pt] $9.0 $  \\[-3pt] $ $ \end{tabular} & 
  \begin{tabular}{l} $0.247339-0.108003i $  \\[-3pt] $0.247178-0.108068i $ \\[-3pt] $0.247179-0.108068i $ \end{tabular} & 
  \begin{tabular}{l} $ $ \\[-3pt] $16 $ \\[-3pt] $ $  \end{tabular} & \begin{tabular}{l} $ $ \\[-3pt] $11 $  \\[-3pt] $ $ \end{tabular} \\[-1pt]
  \hline \hline
  \end{tabular}
  \end{table*}
  \end{center}

  \begin{center}
    \begin{table*}
    \begin{tabular}{p{2cm}p{4.5cm}p{1cm}p{1cm}p{4.5cm}p{1cm}p{1cm}}
    \multicolumn{7}{l}{TABLE II. Fundamental modes of the uncharged cases for the second kind aether black}\\
    \multicolumn{7}{l}{hole with $Q=0 $, $e=0$, $c14=0.2$, obtained by WKB (first line), continued fractions }\\
    \multicolumn{7}{l}{method (second line) and generalized eigenvalue method  (third line).}\\
    \hline \hline
    \begin{tabular}{l} \\ Parameter \\ \hline $c_{13}$ \end{tabular} & 
    \multicolumn{3}{c}{
        \begin{tabular}{p{4.5cm}p{1cm}p{1cm}} 
            \multicolumn{3}{c}{$\ell=0$} \\ \hline
            $\;\;\;\;\;\;\;\;\;\;$ QNM & \multicolumn{2}{c}{Effect \% $\;\;\;\;$}\\ \hline
            $\;\;\;\;\;\;\;\;\;\;\;\;\;$ $\omega$ & $\delta_{\mathrm{Re}}$  & $\delta_{\mathrm{Im}}$
        \end{tabular}
    } &
    \multicolumn{3}{c}{
        \begin{tabular}{p{4.5cm}p{1cm}p{1cm}} 
            \multicolumn{3}{c}{$\ell=1$} \\ \hline
            $\;\;\;\;\;\;\;\;\;\;$ QNM & \multicolumn{2}{c}{Effect \% $\;\;\;\;$}\\ \hline
            $\;\;\;\;\;\;\;\;\;\;\;\;\;$ $\omega$ & $\delta_{\mathrm{Re}}$  & $\delta_{\mathrm{Im}}$
        \end{tabular}
    } \\[3pt] \hline \\[-12pt]
    \begin{tabular}{l} $0.1$ \\[-3pt] $ $ \\[-3pt] $ $ \end{tabular} & 
    \begin{tabular}{l} $0.110678-0.104424i $ \\[-3pt] $0.110455-0.104896i $ \\[-3pt] $0.110455-0.104896i $ \end{tabular} & 
    \begin{tabular}{l} $ $ \\[-3pt] $0 $ \\[-3pt] $ $ \end{tabular} & \begin{tabular}{l} $ $ \\[-3pt] $0 $ \\[-3pt] $ $ \end{tabular} & 
    \begin{tabular}{l} $0.292932-0.097660i $ \\[-3pt] $0.292936-0.097660i $ \\[-3pt] $0.292936-0.097660i $ \end{tabular} & 
    \begin{tabular}{l} $ $ \\[-3pt] $0 $ \\[-3pt] $ $ \end{tabular} & \begin{tabular}{l} $ $ \\[-3pt] $0 $ \\[-3pt] $ $ \end{tabular} \\[-1pt]
    \begin{tabular}{l} $0.25$ \\[-3pt] $ $  \\[-3pt] $ $ \end{tabular} & 
    \begin{tabular}{l} $0.107071-0.103500i $  \\[-3pt] $0.106840-0.103974i $ \\[-3pt] $0.106840-0.103974i $ \end{tabular} & 
    \begin{tabular}{l} $ $ \\[-3pt] $3.3 $  \\[-3pt] $ $ \end{tabular} & \begin{tabular}{l} $ $ \\[-3pt] $0.88 $  \\[-3pt] $ $ \end{tabular} & 
    \begin{tabular}{l} $0.283693-0.096589i $  \\[-3pt] $0.283699-0.096591i $ \\[-3pt] $0.283699-0.096591i $ \end{tabular} & 
    \begin{tabular}{l} $ $ \\[-3pt] $3.2 $  \\[-3pt] $ $ \end{tabular} & \begin{tabular}{l} $ $ \\[-3pt] $1.1 $  \\[-3pt] $ $ \end{tabular} \\[-1pt]
    \begin{tabular}{l} $0.4$ \\[-3pt] $ $  \\[-3pt] $ $\end{tabular} & 
    \begin{tabular}{l} $0.102441-0.101957i $ \\[-3pt] $0.102197-0.102433i $\\[-3pt] $0.102197-0.102433i $ \end{tabular} & 
    \begin{tabular}{l} $ $ \\[-3pt] $7.5 $\\[-3pt] $ $ \end{tabular} & \begin{tabular}{l} $ $ \\[-3pt] $2.3 $\\[-3pt] $ $ \end{tabular} & 
    \begin{tabular}{l} $0.271838-0.094901i $ \\[-3pt] $0.271846-0.094906i $\\[-3pt] $0.271846-0.094906i $ \end{tabular} & 
    \begin{tabular}{l} $ $ \\[-3pt] $7.2 $\\[-3pt] $ $ \end{tabular} & \begin{tabular}{l} $ $ \\[-3pt] $2.8 $ \\[-3pt]  $ $ \end{tabular} \\[-1pt]
    \begin{tabular}{l} $0.55$ \\[-3pt] $ $ \\[-3pt] $ $ \end{tabular} & 
    \begin{tabular}{l} $0.096215-0.099294i $ \\[-3pt] $0.095936-0.099779i $\\[-3pt] $0.095936-0.099779i $ \end{tabular} & 
    \begin{tabular}{l} $ $ \\[-3pt] $13 $ \\[-3pt] $ $\end{tabular} & \begin{tabular}{l} $ $ \\[-3pt] $4.9 $\\[-3pt] $ $ \end{tabular} & 
    \begin{tabular}{l} $0.255829-0.092123i $\\[-3pt] $0.255838-0.092133i $ \\[-3pt] $0.255838-0.092133i $ \end{tabular} & 
    \begin{tabular}{l} $ $ \\[-3pt] $13 $\\[-3pt] $ $ \end{tabular} & \begin{tabular}{l} $ $ \\[-3pt] $5.7 $\\[-3pt] $ $ \end{tabular} \\[-1pt]
    \begin{tabular}{l} $0.70$ \\[-3pt] $ $ \\[-3pt] $ $\end{tabular} & 
    \begin{tabular}{l} $0.087179-0.094317i $\\[-3pt] $0.086794-0.094836i $ \\[-3pt] $0.086794-0.094835i $ \end{tabular} & 
    \begin{tabular}{l} $ $ \\[-3pt] $21 $ \\[-3pt] $ $\end{tabular} & \begin{tabular}{l} $ $ \\[-3pt] $9.6 $ \\[-3pt] $ $\end{tabular} & 
    \begin{tabular}{l} $0.232334-0.087139i $\\[-3pt] $0.232341-0.087156i $ \\[-3pt] $0.232341-0.087156i $ \end{tabular} & 
    \begin{tabular}{l} $ $ \\[-3pt] $21 $\\[-3pt] $ $ \end{tabular} & \begin{tabular}{l} $ $ \\[-3pt] $11 $ \\[-3pt] $ $\end{tabular} \\[-1pt]
    \begin{tabular}{l} $0.85$ \\[-3pt] $ $ \\[-3pt] $ $\end{tabular} & 
    \begin{tabular}{l} $0.071753-0.083125i $ \\[-3pt] $0.071112-0.083732i $\\[-3pt] $0.071107-0.083730i $ \end{tabular} & 
    \begin{tabular}{l} $ $ \\[-3pt] $36 $ \\[-3pt] $ $\end{tabular} & \begin{tabular}{l} $ $ \\[-3pt] $20 $ \\[-3pt] $ $\end{tabular} & 
    \begin{tabular}{l} $0.191571-0.076340i $ \\[-3pt] $0.191576-0.076364i $\\[-3pt] $0.191576-0.076364i $ \end{tabular} & 
    \begin{tabular}{l} $ $ \\[-3pt] $35 $ \\[-3pt] $ $\end{tabular} & \begin{tabular}{l} $ $ \\[-3pt] $22 $ \\[-3pt] $ $\end{tabular} \\[-1pt]
    \hline \hline
    \end{tabular}
    \end{table*}
    \end{center}

First we analyze the accuracy of the continued fraction method and the generalized eigenvalue method by comparing the results among these methods. In Table I and Table II, we show the results obtained by Churilova \cite{Churilova2020} through the sixth-order WKB formula at the first line, the continued fraction method at the second line and the generalized eigenvalue method at the third line. We calculate the extra values for $l=1$ to make a better comparison. For the aether cases, the results of these methods turn out to be in good agreement with each other for the different values of $c_{13}$, especially for the continued fraction method and the generalized eigenvalue method.

The percentages of relative effect are placed to the right of the quasi-normal frequencies for each $c_{13}$ respectively. Previously there are the relative effects obtained by third-order WKB \cite{Churilova2020}. We use more accurate results to demonstrate the aether effect on quasi-normal frequencies. For the first kind aether black hole, the effects of large $c_{13}$, which are compared to the Schwarzschild cases, are $26$ for the real part of the value and $9.0$ for the imaginary part with $l=0$. These are smaller than the effect of third-order WKB. The corresponding effects are following: $16$ and $11$ for the first kind aether black hole with $l=1$, $36$ and $20$ for the second kind aether black hole with $l=0$, $35$ and $22$ for the second kind aether black hole with $l=1$. The relative effects for the second kind aether black hole are larger than those for the first kind, which is contrary to the results of third-order WKB for $l=0$ presented in \cite{Churilova2020}.

\subsection{Spectrum of charged black hole}

    \begin{center}
      \begin{table*}
      \begin{tabular}{p{2cm}p{4.5cm}p{1cm}p{1cm}p{4.5cm}p{1cm}p{1cm}}
          \multicolumn{7}{l}{TABLE III. Fundamental modes of the charged cases for the first kind aether black hole}\\
          \multicolumn{7}{l}{ with $Q=0.1$, $e=0.1$, obtained by WKB (first line), continued fraction method (second }\\
          \multicolumn{7}{l}{line) and generalized eigenvalue method  (third line).}\\
      \hline \hline
      \begin{tabular}{l} \\ Parameter \\ \hline $c_{13}$ \end{tabular} & 
      \multicolumn{3}{c}{
          \begin{tabular}{p{4.5cm}p{1cm}p{1cm}} 
              \multicolumn{3}{c}{$\ell=1$} \\ \hline
              $\;\;\;\;\;\;\;\;\;\;$ QNM & \multicolumn{2}{c}{Effect \% $\;\;\;\;$}\\ \hline
              $\;\;\;\;\;\;\;\;\;\;\;\;\;$ $\omega$ & $\delta_{\mathrm{Re}}$  & $\delta_{\mathrm{Im}}$
          \end{tabular}
      } &
      \multicolumn{3}{c}{
          \begin{tabular}{p{4.5cm}p{1cm}p{1cm}} 
              \multicolumn{3}{c}{$\ell=2$} \\ \hline
              $\;\;\;\;\;\;\;\;\;\;$ QNM & \multicolumn{2}{c}{Effect \% $\;\;\;\;$}\\ \hline
              $\;\;\;\;\;\;\;\;\;\;\;\;\;$ $\omega$ & $\delta_{\mathrm{Re}}$  & $\delta_{\mathrm{Im}}$
          \end{tabular}
      } \\[3pt] \hline \\[-12pt]
      \begin{tabular}{l} $0$ \\[-3pt] $ $ \\[-3pt] $ $ \end{tabular} & 
      \begin{tabular}{l} $0.298515-0.098187i$ \\[-3pt] $0.298408-0.098209i $ \\[-3pt] $0.296162-0.098874i $   \end{tabular} & 
      \begin{tabular}{l} $ $ \\[-3pt] $0$ \\[-3pt]$ $ \end{tabular} & \begin{tabular}{l} $ $ \\[-3pt] $0$ \\[-3pt]$ $ \end{tabular} & 
      \begin{tabular}{l} $0.490396-0.097187i $ \\[-3pt]$0.490334-0.097184i $ \\[-3pt] $0.491114-0.096766i $  \end{tabular} & 
      \begin{tabular}{l} $ $ \\[-3pt] $0$ \\[-3pt]$ $ \end{tabular} & \begin{tabular}{l} $ $ \\[-3pt] $0$\\[-3pt]$ $  \end{tabular} \\[-1pt]
      \begin{tabular}{l} $0.1$ \\[-3pt] $ $ \\[-3pt] $ $ \end{tabular} & 
      \begin{tabular}{l} $0.297372-0.098916i$ \\[-3pt] $0.297274-0.098931i $ \\[-3pt] $0.295114-0.098816i $   \end{tabular} & 
      \begin{tabular}{l} $ $ \\[-3pt] $0.35$ \\[-3pt]$ $ \end{tabular} & \begin{tabular}{l} $ $ \\[-3pt] $0.06$ \\[-3pt]$ $ \end{tabular} & 
      \begin{tabular}{l} $0.488661-0.097879i $ \\[-3pt]$0.488615-0.097866i $ \\[-3pt] $0.489344-0.097633i $  \end{tabular} & 
      \begin{tabular}{l} $ $ \\[-3pt] $0.36$ \\[-3pt]$ $ \end{tabular} & \begin{tabular}{l} $ $ \\[-3pt] $0.9$\\[-3pt]$ $  \end{tabular} \\[-1pt]
      \begin{tabular}{l} $0.2$ \\[-3pt] $ $ \\[-3pt] $ $ \end{tabular} & 
      \begin{tabular}{l} $0.296012-0.099719i $\\[-3pt] $0.295894-0.099743i $ \\[-3pt] $0.293759-0.099197i $   \end{tabular} & 
      \begin{tabular}{l} $ $ \\[-3pt] $0.81 $ \\[-3pt] $ $  \end{tabular} & \begin{tabular}{l} $ $ \\[-3pt] $0.33 $  \\[-3pt] $ $ \end{tabular} & 
      \begin{tabular}{l} $0.486593-0.098624i $ \\[-3pt] $0.486539-0.098639i $ \\[-3pt] $0.487324-0.098578i $   \end{tabular} & 
      \begin{tabular}{l} $ $ \\[-3pt] $0.77 $ \\[-3pt] $ $  \end{tabular} & \begin{tabular}{l} $ $ \\[-3pt] $1.9 $  \\[-3pt] $ $ \end{tabular} \\[-1pt]
      \begin{tabular}{l} $0.3$ \\[-3pt] $ $ \\[-3pt] $ $  \end{tabular} & 
      \begin{tabular}{l} $0.294295-0.100666i $ \\[-3pt] $0.294185-0.100660i $ \\[-3pt] $0.292192-0.099639i $   \end{tabular} & 
      \begin{tabular}{l} $ $ \\[-3pt] $1.3 $ \\[-3pt] $ $  \end{tabular} & \begin{tabular}{l} $ $ \\[-3pt] $0.77 $ \\[-3pt] $ $  \end{tabular} & 
      \begin{tabular}{l} $0.484009-0.099517i $\\[-3pt] $0.483982-0.099520i $ \\[-3pt] $0.484797-0.099676i $  \end{tabular} & 
      \begin{tabular}{l} $ $ \\[-3pt] $1.3 $ \\[-3pt] $ $  \end{tabular} & \begin{tabular}{l} $ $ \\[-3pt] $3.0 $ \\[-3pt] $ $  \end{tabular} \\[-1pt]
      \begin{tabular}{l} $0.4$ \\[-3pt] $ $ \\[-3pt] $ $  \end{tabular} & 
      \begin{tabular}{l} $0.292124-0.101662i $  \\[-3pt] $0.292015-0.101706i $ \\[-3pt] $0.290327-0.100148i $  \end{tabular} & 
      \begin{tabular}{l} $ $ \\[-3pt] $2.0 $ \\[-3pt] $ $  \end{tabular} & \begin{tabular}{l} $ $ \\[-3pt] $1.3 $ \\[-3pt] $ $  \end{tabular} & 
      \begin{tabular}{l} $0.480820-0.100522i $\\[-3pt] $0.480752-0.100534i $ \\[-3pt] $0.481489-0.100966i $   \end{tabular} & 
      \begin{tabular}{l} $ $ \\[-3pt] $2.0 $ \\[-3pt] $ $  \end{tabular} & \begin{tabular}{l} $ $ \\[-3pt] $4.3 $ \\[-3pt] $ $  \end{tabular} \\[-1pt]
      \begin{tabular}{l} $0.5$ \\[-3pt] $ $ \\[-3pt] $ $  \end{tabular} & 
      \begin{tabular}{l} $0.289291-0.102890i $\\[-3pt] $0.289173-0.102908i $ \\[-3pt] $0.288042-0.100862i $   \end{tabular} & 
      \begin{tabular}{l} $ $ \\[-3pt] $2.7 $ \\[-3pt] $ $  \end{tabular} & \begin{tabular}{l} $ $ \\[-3pt] $2.0 $  \\[-3pt] $ $ \end{tabular} & 
      \begin{tabular}{l} $0.476578-0.101691i $\\[-3pt] $0.476537-0.101712i $ \\[-3pt] $0.477082-0.102462i $   \end{tabular} & 
      \begin{tabular}{l} $ $ \\[-3pt] $1.7 $ \\[-3pt] $ $  \end{tabular} & \begin{tabular}{l} $ $ \\[-3pt] $5.2 $ \\[-3pt] $ $  \end{tabular} \\[-1pt]
      \begin{tabular}{l} $0.6$ \\[-3pt] $ $ \\[-3pt] $ $  \end{tabular} & 
      \begin{tabular}{l} $0.285391-0.104261i $ \\[-3pt] $0.285288-0.104296i $ \\[-3pt] $0.285072-0.101867i $   \end{tabular} & 
      \begin{tabular}{l} $ $ \\[-3pt] $3.7 $ \\[-3pt] $ $  \end{tabular} & \begin{tabular}{l} $ $ \\[-3pt] $3.0 $  \\[-3pt] $ $ \end{tabular} & 
      \begin{tabular}{l} $0.470854-0.103101i $ \\[-3pt] $0.470785-0.103092i $ \\[-3pt] $0.470848-0.104048i $   \end{tabular} & 
      \begin{tabular}{l} $ $ \\[-3pt] $4.1 $ \\[-3pt] $ $  \end{tabular} & \begin{tabular}{l} $ $ \\[-3pt] $7.5 $  \\[-3pt] $ $ \end{tabular} \\[-1pt]            
      \begin{tabular}{l} $0.7$ \\[-3pt] $ $ \\[-3pt] $ $  \end{tabular} & 
      \begin{tabular}{l} $0.279753-0.105835i $ \\[-3pt] $0.279630-0.105895i $ \\[-3pt] $0.280776-0.103629i $   \end{tabular} & 
      \begin{tabular}{l} $ $ \\[-3pt] $5.2 $ \\[-3pt] $ $  \end{tabular} & \begin{tabular}{l} $ $ \\[-3pt] $4.8 $  \\[-3pt] $ $ \end{tabular} & 
      \begin{tabular}{l} $0.462483-0.104714i $ \\[-3pt] $0.462401-0.104713i $ \\[-3pt] $0.461956-0.105532i $   \end{tabular} & 
      \begin{tabular}{l} $ $ \\[-3pt] $6.0 $ \\[-3pt] $ $  \end{tabular} & \begin{tabular}{l} $ $ \\[-3pt] $9.1 $  \\[-3pt] $ $ \end{tabular} \\[-1pt]
      \begin{tabular}{l} $0.8$ \\[-3pt] $ $  \\[-3pt] $ $ \end{tabular} & 
      \begin{tabular}{l} $0.270712-0.107617i $ \\[-3pt] $0.270475-0.107660i $ \\[-3pt] $0.273199-0.106902i $   \end{tabular} & 
      \begin{tabular}{l} $ $ \\[-3pt] $7.8 $ \\[-3pt] $ $  \end{tabular} & \begin{tabular}{l} $ $ \\[-3pt] $8.1 $  \\[-3pt] $ $ \end{tabular} & 
      \begin{tabular}{l} $0.448836-0.106554i $ \\[-3pt] $0.448771-0.106560i $ \\[-3pt] $0.447697-0.106282i $   \end{tabular} & 
      \begin{tabular}{l} $ $ \\[-3pt] $8.8 $ \\[-3pt] $ $  \end{tabular} & \begin{tabular}{l} $ $ \\[-3pt] $9.8 $  \\[-3pt] $ $ \end{tabular} \\[-1pt]            
      \begin{tabular}{l} $0.9$ \\[-3pt] $ $  \\[-3pt] $ $ \end{tabular} & 
      \begin{tabular}{l} $0.252367-0.108843i $ \\[-3pt] $0.252062-0.108975i $ \\[-3pt] $0.252412-0.112625i $   \end{tabular} & 
      \begin{tabular}{l} $ $ \\[-3pt] $15 $  \\[-3pt] $ $ \end{tabular} & \begin{tabular}{l} $ $ \\[-3pt] $14 $ \\[-3pt] $ $  \end{tabular} & 
      \begin{tabular}{l} $0.421103-0.108100i $ \\[-3pt] $0.421016-0.108109i $ \\[-3pt] $0.421709-0.107267i $   \end{tabular} & 
      \begin{tabular}{l} $ $ \\[-3pt] $14 $  \\[-3pt] $ $ \end{tabular} & \begin{tabular}{l} $ $ \\[-3pt] $11 $  \\[-3pt] $ $ \end{tabular} \\[-1pt]
      \hline \hline
      \end{tabular}
      \end{table*}
      \end{center}

      \begin{center}
        \begin{table*}
        \begin{tabular}{p{2cm}p{4.5cm}p{1cm}p{1cm}p{4.5cm}p{1cm}p{1cm}}
        \multicolumn{7}{l}{TABLE IV. Fundamental modes of the charged cases for the second kind aether black }\\
        \multicolumn{7}{l}{hole with $Q=0.1$, $e=0.1$, $c14=0.2$, obtained by WKB (first line), continued fractions}\\
        \multicolumn{7}{l}{method (second line) and generalized eigenvalue method  (third line).}\\
        \hline \hline
        \begin{tabular}{l} \\ Parameter \\ \hline $c_{13}$ \end{tabular} & 
        \multicolumn{3}{c}{
            \begin{tabular}{p{4.5cm}p{1cm}p{1cm}} 
                \multicolumn{3}{c}{$\ell=1$} \\ \hline
                $\;\;\;\;\;\;\;\;\;\;$ QNM & \multicolumn{2}{c}{Effect \% $\;\;\;\;$}\\ \hline
                $\;\;\;\;\;\;\;\;\;\;\;\;\;$ $\omega$ & $\delta_{\mathrm{Re}}$  & $\delta_{\mathrm{Im}}$
            \end{tabular}
        } &
        \multicolumn{3}{c}{
            \begin{tabular}{p{4.5cm}p{1cm}p{1cm}} 
                \multicolumn{3}{c}{$\ell=2$} \\ \hline
                $\;\;\;\;\;\;\;\;\;\;$ QNM & \multicolumn{2}{c}{Effect \% $\;\;\;\;$}\\ \hline
                $\;\;\;\;\;\;\;\;\;\;\;\;\;$ $\omega$ & $\delta_{\mathrm{Re}}$  & $\delta_{\mathrm{Im}}$
            \end{tabular}
        } \\[3pt] \hline \\[-12pt]
        \begin{tabular}{l} $0.15$ \\[-3pt] $ $ \\[-3pt] $ $  \end{tabular} & 
        \begin{tabular}{l} $0.295850-0.097916i $ \\[-3pt] $0.295782-0.097947i $ \\[-3pt] $0.293726-0.097687i $   \end{tabular} & 
        \begin{tabular}{l} $ $ \\[-3pt] $0.8 $ \\[-3pt] $ $  \end{tabular} & \begin{tabular}{l} $ $ \\[-3pt] $1.2 $  \\[-3pt] $ $ \end{tabular} & 
        \begin{tabular}{l} $0.486082-0.096894i $\\[-3pt] $0.486059-0.096914i $ \\[-3pt] $0.486787-0.096749i $   \end{tabular} & 
        \begin{tabular}{l} $ $ \\[-3pt] $0.88 $ \\[-3pt] $ $  \end{tabular} & \begin{tabular}{l} $ $ \\[-3pt] $0.02 $  \\[-3pt] $ $ \end{tabular} \\[-1pt]
        \begin{tabular}{l} $0.25$ \\[-3pt] $ $ \\[-3pt] $ $  \end{tabular} & 
        \begin{tabular}{l} $0.289457-0.097198i $ \\[-3pt] $0.289374-0.097230i $ \\[-3pt] $0.287875-0.095979i $   \end{tabular} & 
        \begin{tabular}{l} $ $ \\[-3pt] $2.8 $ \\[-3pt] $ $  \end{tabular} & \begin{tabular}{l} $ $ \\[-3pt] $2.9 $ \\[-3pt] $ $  \end{tabular} & 
        \begin{tabular}{l} $0.475662-0.096160i $ \\[-3pt] $0.475623-0.096180i $ \\[-3pt] $0.476295-0.096520i $   \end{tabular} & 
        \begin{tabular}{l} $ $ \\[-3pt] $3.0 $  \\[-3pt] $ $ \end{tabular} & \begin{tabular}{l} $ $ \\[-3pt] $0.26 $ \\[-3pt] $ $  \end{tabular} \\[-1pt]
        \begin{tabular}{l} $0.35$ \\[-3pt] $ $ \\[-3pt] $ $  \end{tabular} & 
        \begin{tabular}{l} $0.281878-0.096256i $  \\[-3pt] $0.281804-0.096249i $ \\[-3pt] $0.281364-0.094458i $  \end{tabular} & 
        \begin{tabular}{l} $ $ \\[-3pt] $5.0 $ \\[-3pt] $ $  \end{tabular} & \begin{tabular}{l} $ $ \\[-3pt] $4.5 $  \\[-3pt] $ $ \end{tabular} & 
        \begin{tabular}{l} $0.463344-0.095184i $ \\[-3pt] $0.463294-0.095183i $ \\[-3pt] $0.463508-0.095919i $   \end{tabular} & 
        \begin{tabular}{l} $ $ \\[-3pt] $5.6 $ \\[-3pt] $ $  \end{tabular} & \begin{tabular}{l} $ $ \\[-3pt] $0.88 $  \\[-3pt] $ $ \end{tabular} \\[-1pt]
        \begin{tabular}{l} $0.45$ \\[-3pt] $ $  \\[-3pt] $ $ \end{tabular} & 
        \begin{tabular}{l} $0.272759-0.094857i $ \\[-3pt] $0.272674-0.094888i $ \\[-3pt] $0.273514-0.093325i $   \end{tabular} & 
        \begin{tabular}{l} $ $ \\[-3pt] $7.6 $  \\[-3pt] $ $ \end{tabular} & \begin{tabular}{l} $ $ \\[-3pt] $5.6 $  \\[-3pt] $ $ \end{tabular} & 
        \begin{tabular}{l} $0.448459-0.093790i $ \\[-3pt] $0.448418-0.093807i $ \\[-3pt] $0.447911-0.094392i $   \end{tabular} & 
        \begin{tabular}{l} $ $ \\[-3pt] $8.8 $  \\[-3pt] $ $ \end{tabular} & \begin{tabular}{l} $ $ \\[-3pt] $2.5 $  \\[-3pt] $ $ \end{tabular} \\[-1pt]
        \begin{tabular}{l} $0.55$ \\[-3pt] $ $  \\[-3pt] $ $ \end{tabular} & 
        \begin{tabular}{l} $0.261437-0.092922i $\\[-3pt] $0.261352-0.092950i $ \\[-3pt] $0.263099-0.092647i $   \end{tabular} & 
        \begin{tabular}{l} $ $ \\[-3pt] $11 $  \\[-3pt] $ $ \end{tabular} & \begin{tabular}{l} $ $ \\[-3pt] $6.3 $  \\[-3pt] $ $ \end{tabular} & 
        \begin{tabular}{l} $0.430002-0.091843i $ \\[-3pt] $0.429962-0.091855i $ \\[-3pt] $0.429258-0.091596i $   \end{tabular} & 
        \begin{tabular}{l} $ $ \\[-3pt] $13 $  \\[-3pt] $ $ \end{tabular} & \begin{tabular}{l} $ $ \\[-3pt] $5.3 $  \\[-3pt] $ $ \end{tabular} \\[-1pt]
        \begin{tabular}{l} $0.65$ \\[-3pt] $ $ \\[-3pt] $ $  \end{tabular} & 
        \begin{tabular}{l} $0.246798-0.090055i $ \\[-3pt] $0.246753-0.090077i $ \\[-3pt] $0.247651-0.091666i $   \end{tabular} & 
        \begin{tabular}{l} $ $ \\[-3pt] $16 $ \\[-3pt] $ $ \end{tabular} & \begin{tabular}{l} $ $ \\[-3pt] $7.3 $  \\[-3pt] $ $ \end{tabular} & 
        \begin{tabular}{l} $0.406182-0.088968i $ \\[-3pt] $0.406144-0.088974i $ \\[-3pt] $0.406323-0.088281i $   \end{tabular} & 
        \begin{tabular}{l} $ $ \\[-3pt] $17 $  \\[-3pt] $ $ \end{tabular} & \begin{tabular}{l} $ $ \\[-3pt] $8.8 $  \\[-3pt] $ $ \end{tabular} \\[-1pt]
        \begin{tabular}{l} $0.75$ \\[-3pt] $ $ \\[-3pt] $ $  \end{tabular} & 
        \begin{tabular}{l} $0.226836-0.085443i $ \\[-3pt] $0.226766-0.085531i $ \\[-3pt] $0.225215-0.086017i $   \end{tabular} & 
        \begin{tabular}{l} $ $ \\[-3pt] $24 $ \\[-3pt] $ $  \end{tabular} & \begin{tabular}{l} $ $ \\[-3pt] $13 $ \\[-3pt] $ $  \end{tabular} & 
        \begin{tabular}{l} $0.373518-0.084436i $ \\[-3pt] $0.373498-0.084431i $ \\[-3pt] $0.374071-0.084820i $  \end{tabular} & 
        \begin{tabular}{l} $ $ \\[-3pt] $24 $ \\[-3pt] $ $  \end{tabular} & \begin{tabular}{l} $ $ \\[-3pt] $12 $ \\[-3pt] $ $  \end{tabular} \\[-1pt]
        \begin{tabular}{l} $0.85$ \\[-3pt] $ $ \\[-3pt] $ $  \end{tabular} & 
        \begin{tabular}{l} $0.196390-0.077350i $ \\[-3pt] $0.196333-0.077411i $ \\[-3pt] $0.197072-0.076325i $   \end{tabular} & 
        \begin{tabular}{l} $ $ \\[-3pt] $33 $  \\[-3pt] $ $ \end{tabular} & \begin{tabular}{l} $ $ \\[-3pt] $23 $  \\[-3pt] $ $ \end{tabular} & 
        \begin{tabular}{l} $0.323737-0.076333i $ \\[-3pt] $0.323698-0.076348i $\\[-3pt] $0.323221-0.076001i $   \end{tabular} & 
        \begin{tabular}{l} $ $ \\[-3pt] $34 $ \\[-3pt] $ $  \end{tabular} & \begin{tabular}{l} $ $ \\[-3pt] $21 $  \\[-3pt] $ $ \end{tabular} \\[-1pt]
        \begin{tabular}{l} $0.95$ \\[-3pt] $ $ \\[-3pt] $ $ \end{tabular} & 
        \begin{tabular}{l} $0.135844-0.057626i $ \\[-3pt] $0.135865-0.057691i $ \\[-3pt] $0.135386-0.057071i $   \end{tabular} & 
        \begin{tabular}{l} $ $ \\[-3pt] $54 $ \\[-3pt] $ $  \end{tabular} & \begin{tabular}{l} $ $ \\[-3pt] $42 $  \\[-3pt] $ $ \end{tabular} & 
        \begin{tabular}{l} $0.224452-0.056835i $ \\[-3pt] $0.224429-0.056803i $ \\[-3pt] $0.224014-0.056833i $   \end{tabular} & 
        \begin{tabular}{l} $ $ \\[-3pt] $54 $ \\[-3pt] $ $  \end{tabular} & \begin{tabular}{l} $ $ \\[-3pt] $41 $  \\[-3pt] $ $ \end{tabular} \\[-1pt]
        \hline \hline
        \end{tabular}
        \end{table*}
        \end{center}

After that, We go on to the extension of the cases above, the charged scalar perturbations. The results of modes are presented in Table III and Table IV for the first kind and the second kind aether black hole respectively. The results are placed one under the other as the uncharged cases. The values of the quasi-normal modes for the Reissner-Nordström black hole ($c_{13}=0$) presented in Table III are used to calculate the relative effect. Obviously, the results of the WKB method are closer to the continued fraction method than the generalized eigenvalue method. Simply increasing the resolution does not improve the accuracy of the results of the generalized eigenvalue method, resulting in a difference between it and the continued fraction method. In general, the continued fraction method provides the most accurate results compared to the other numerical methods.

The relative effects for charged scalar perturbations are similar to the uncharged case, which is, the effects for the second kind aether black hole are generally larger than the first kind. We can see that the effect of large $c_{13}$ can even exceed $50\%$ in Table IV.
        
\subsection{Effects of charge Q on the quasi-normal modes}

\begin{figure}[htbp]
  \centering
  \includegraphics[width = 0.4\textwidth]{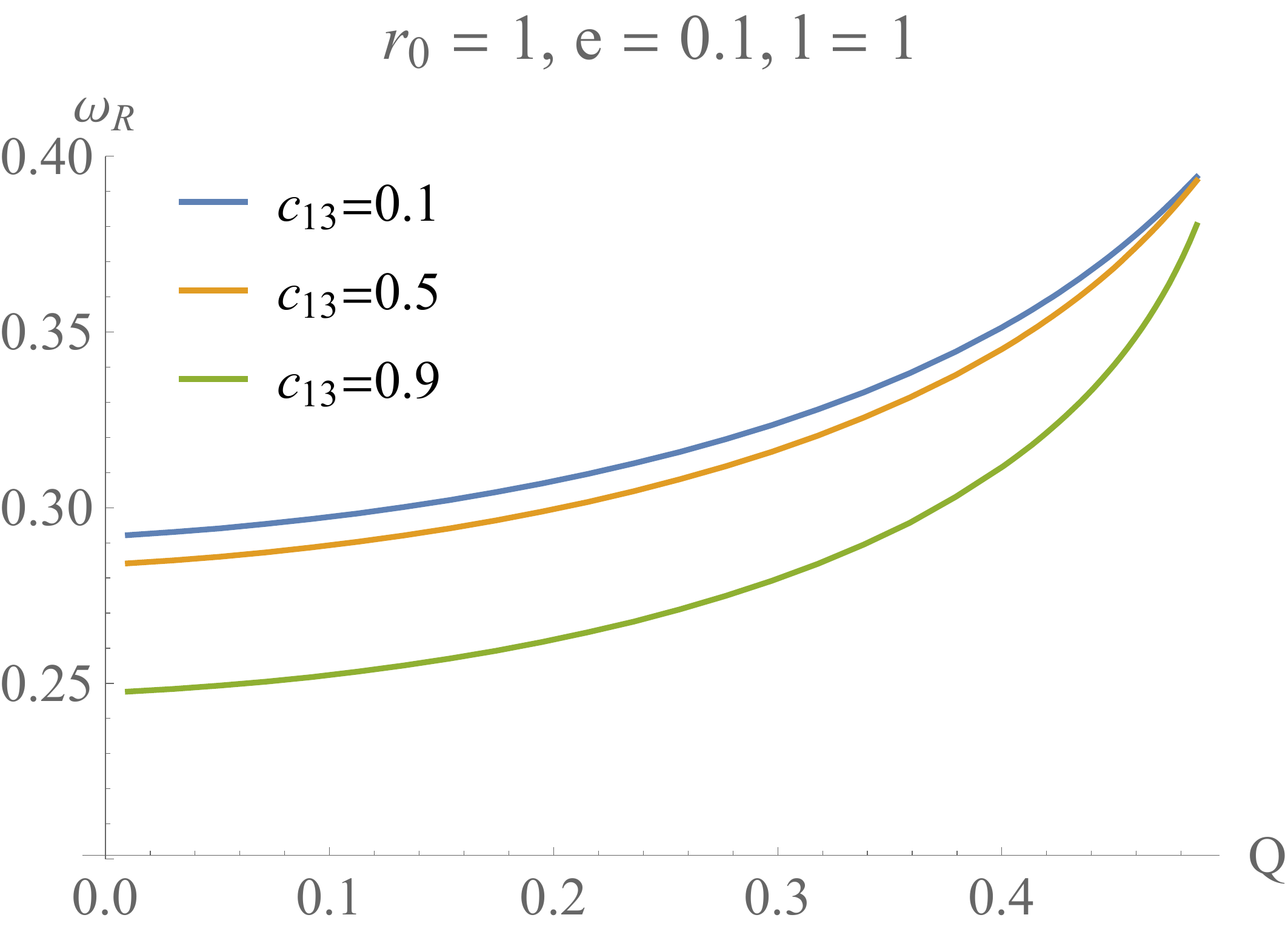}
  \includegraphics[width = 0.4\textwidth]{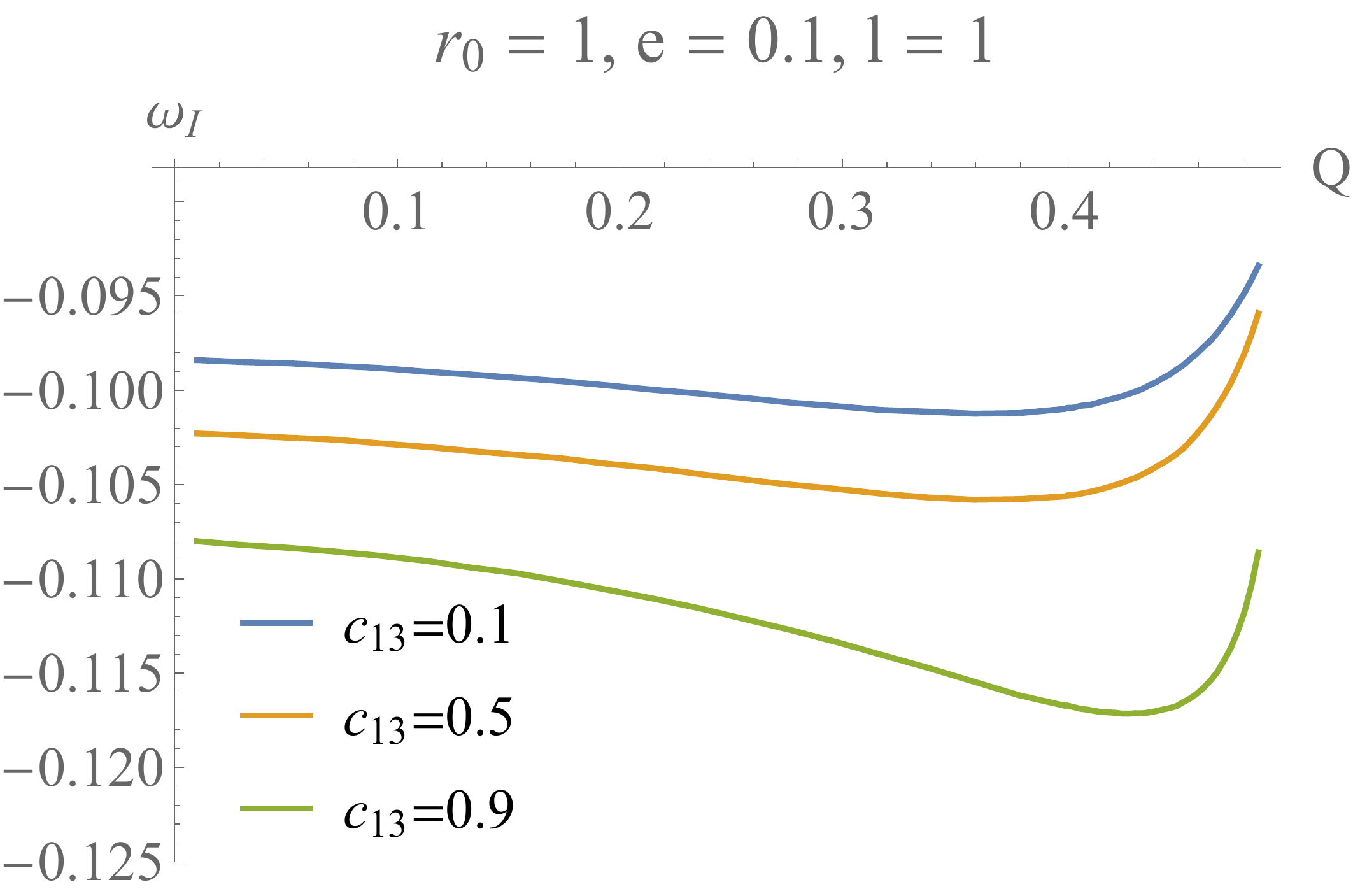}
  \caption{The fundamental modes of charged scalar perturbation vs $Q$ at $r_{0}=1,\,e=0.1,\,l=1$ for the first kind aether black hole.}\label{fig:CS 1 vsQ l=1}
\end{figure}

Then We demonstrate the modes $\omega$ vs $Q$ in Fig. \ref{fig:CS 1 vsQ l=1} with fixed $r_{0}=2, e=0.1, l=1$ for the first kind aether black hole. The real part of the fundamental modes increases with $Q$ monotonically. However, the imaginary part of these frequencies first decreases with $Q$ and then increases with $Q$. Both the real and imaginary part of frequencies get smaller as $c_{13}$ gets larger, which consist of Table II.

The fundamental modes of the second kind aether black hole are following. First of all, we focus on the allowable parameters range for the second kind aether black hole which is very different from the previous cases, because of the parameter constraints (\ref{eq:constraint for c123=0}). These three parameters ($Q, c_{13}, c_{14}$) are intertwined, so we have to decide the parameter range according to which parameter we are varying.

\begin{figure}[htbp]
  \centering
  \includegraphics[width = 0.5\textwidth]{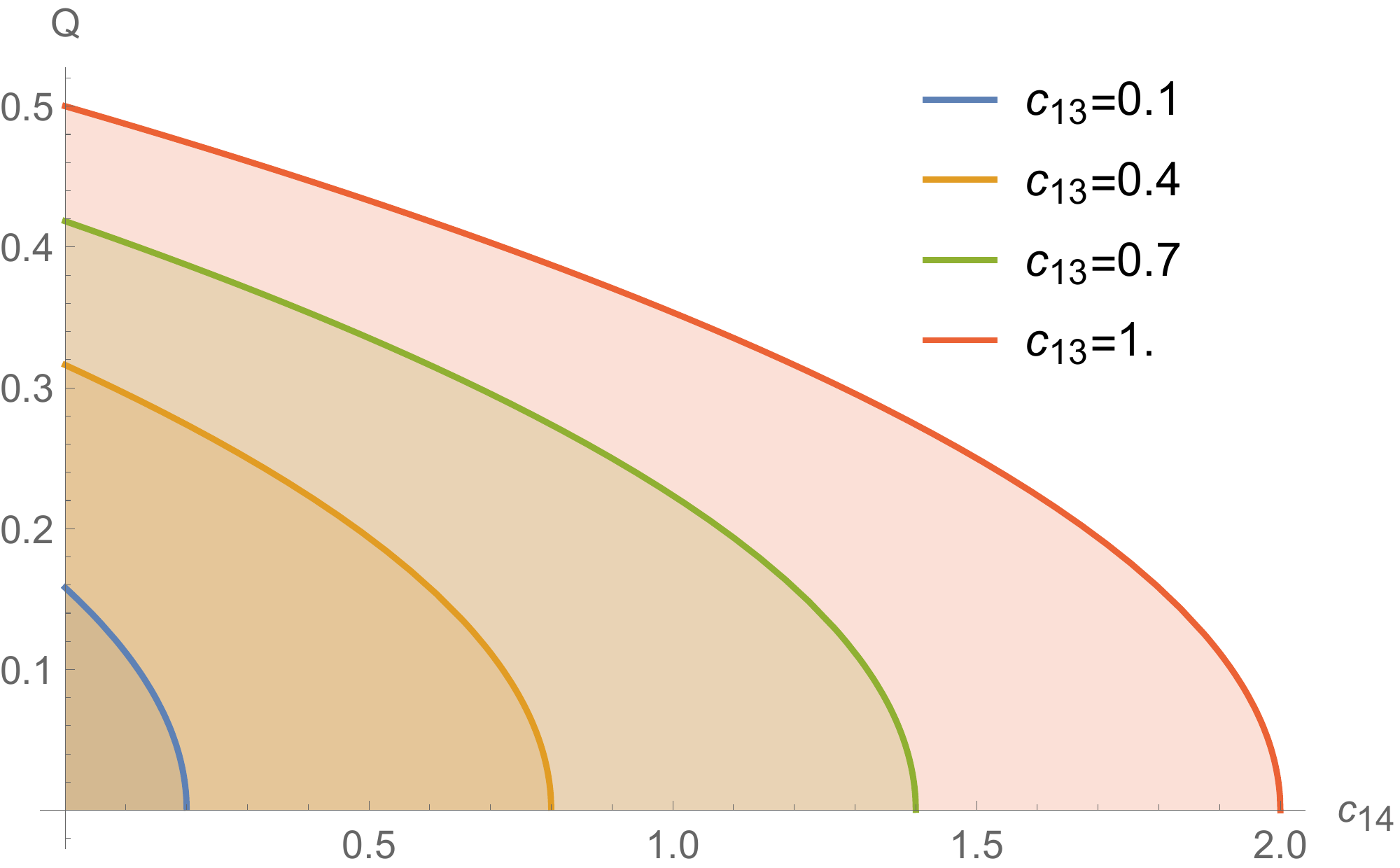}
  \caption{The allowable range of $Q$ for the second kind aether black hole.}\label{fig:CS 2 region Q}
\end{figure}

\begin{figure}[htbp]
  \centering
  \includegraphics[width = 0.4\textwidth]{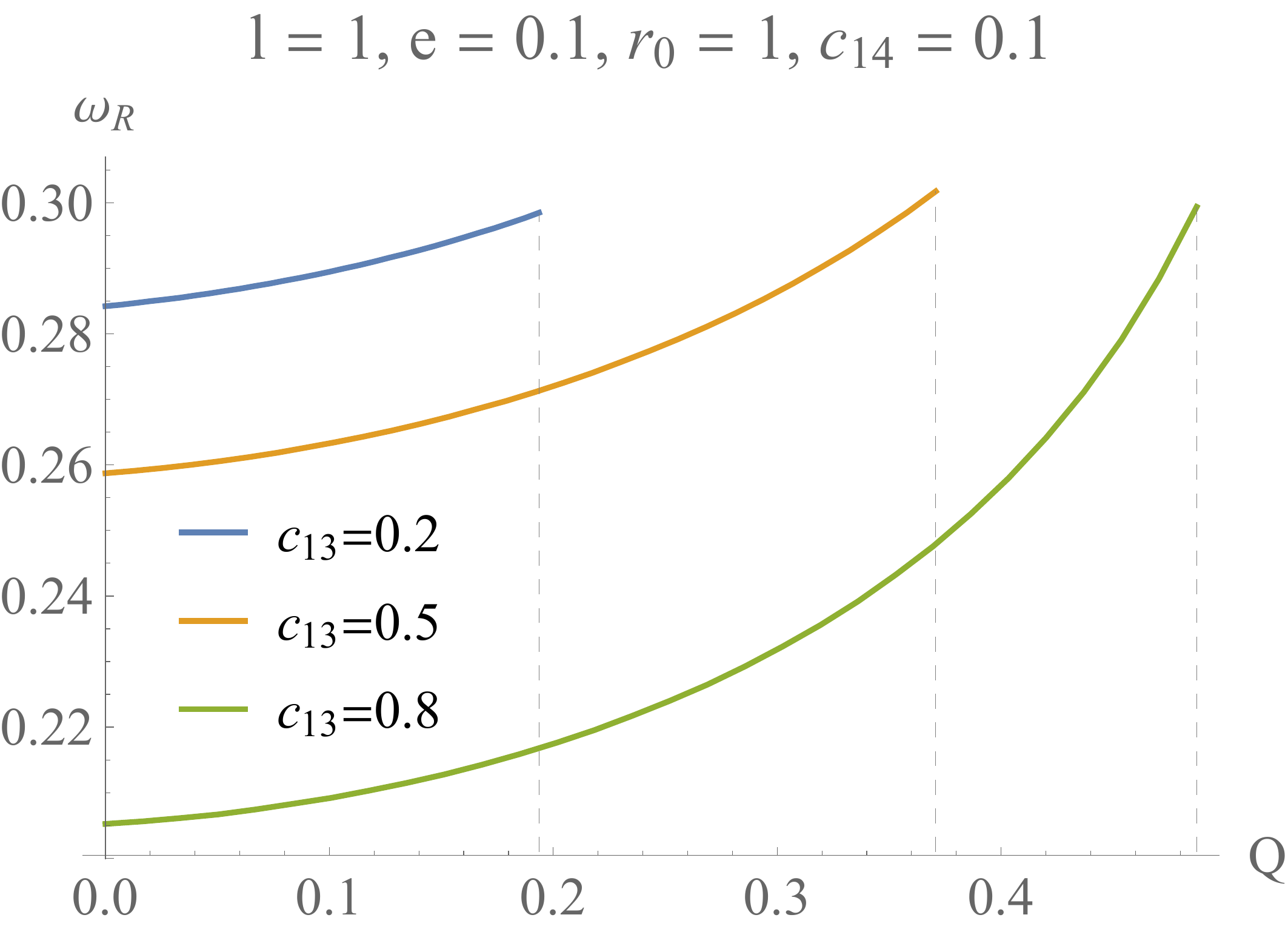}
  \includegraphics[width = 0.4\textwidth]{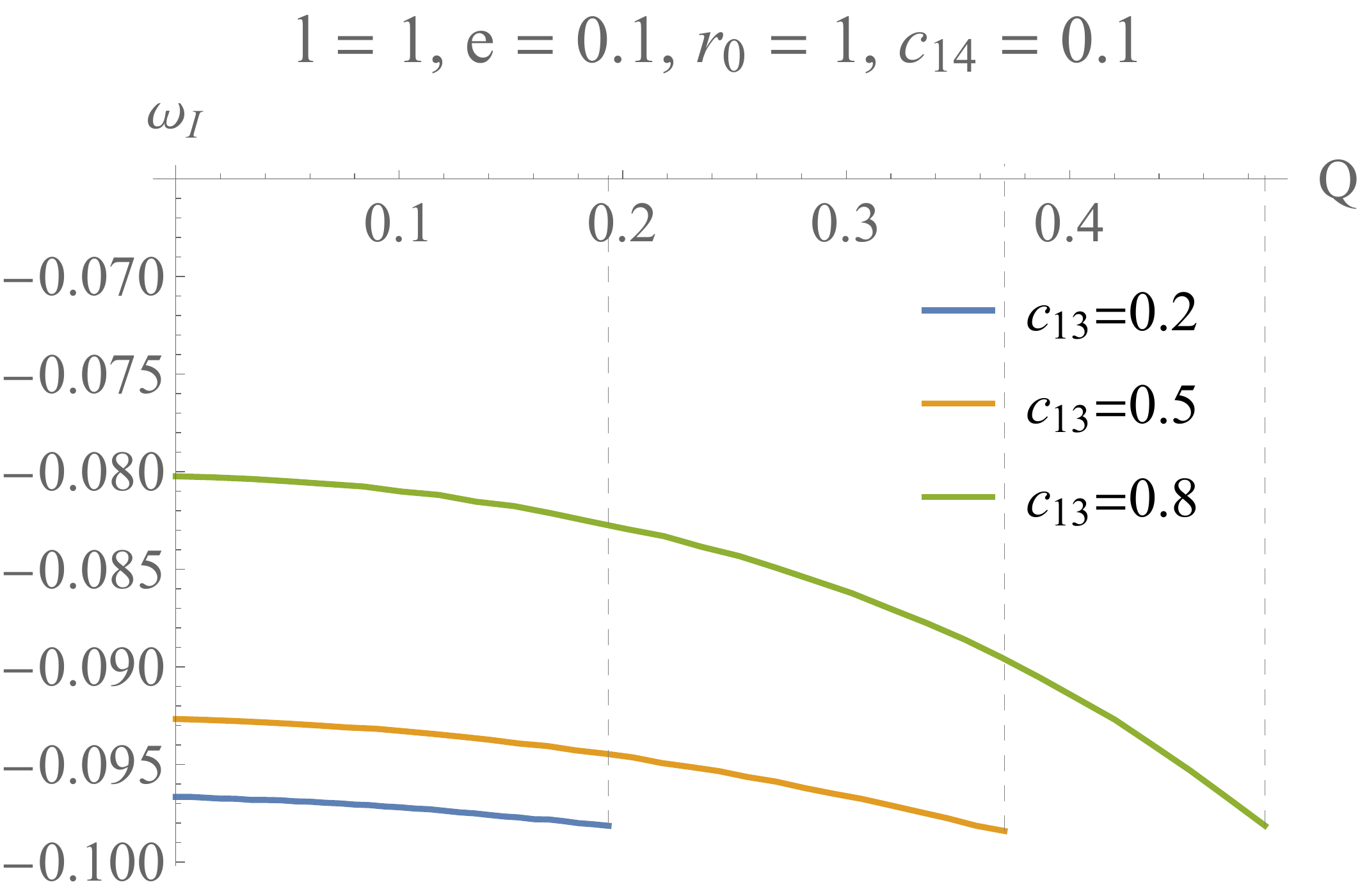}
  \caption{The left and the right plot are $\omega_{\textrm{R}}$ and $\omega_{\textrm{I}}$ of the fundamental modes vs $Q$ at $l=1,\,e=0.1,\,r_{0}=2,\,c_{14}=0.1.$}\label{fig:CS 2 vsQ l=1}
\end{figure}

First we demonstrate the fundamental modes with different $Q$ and Fig. \ref{fig:CS 2 region Q} shows the allowable range of $Q$ obtained by constraints (\ref{eq:constraint for c123=0}), where different curves denote the different choice of $c_{13}$. This plot reveals that the allowable range of $Q$ decreases with $c_{14}$ and even goes to zero as $c_{14}$ goes to its extreme values. In order to show more behaviors of the quasi-normal modes, we always choose the value of fixed parameters $c_{14}$ which can lead to a larger parameter range of the varying parameter $Q$. In this case, small $c_{14}$ generates a large range of $Q$, so we fix $c_{14}=0.1$ and the results are shown by Fig. \ref{fig:CS 2 vsQ l=1}. These plots show that the real part $\omega_R$ increases with $Q$ and the imaginary part $\omega_I$ decreases with $Q$. The larger $c_{13}$ leads to smaller $\omega_{R}$ and larger $\omega_{I}$.

\begin{figure}[htbp]
  \centering
  \includegraphics[width = 0.4\textwidth]{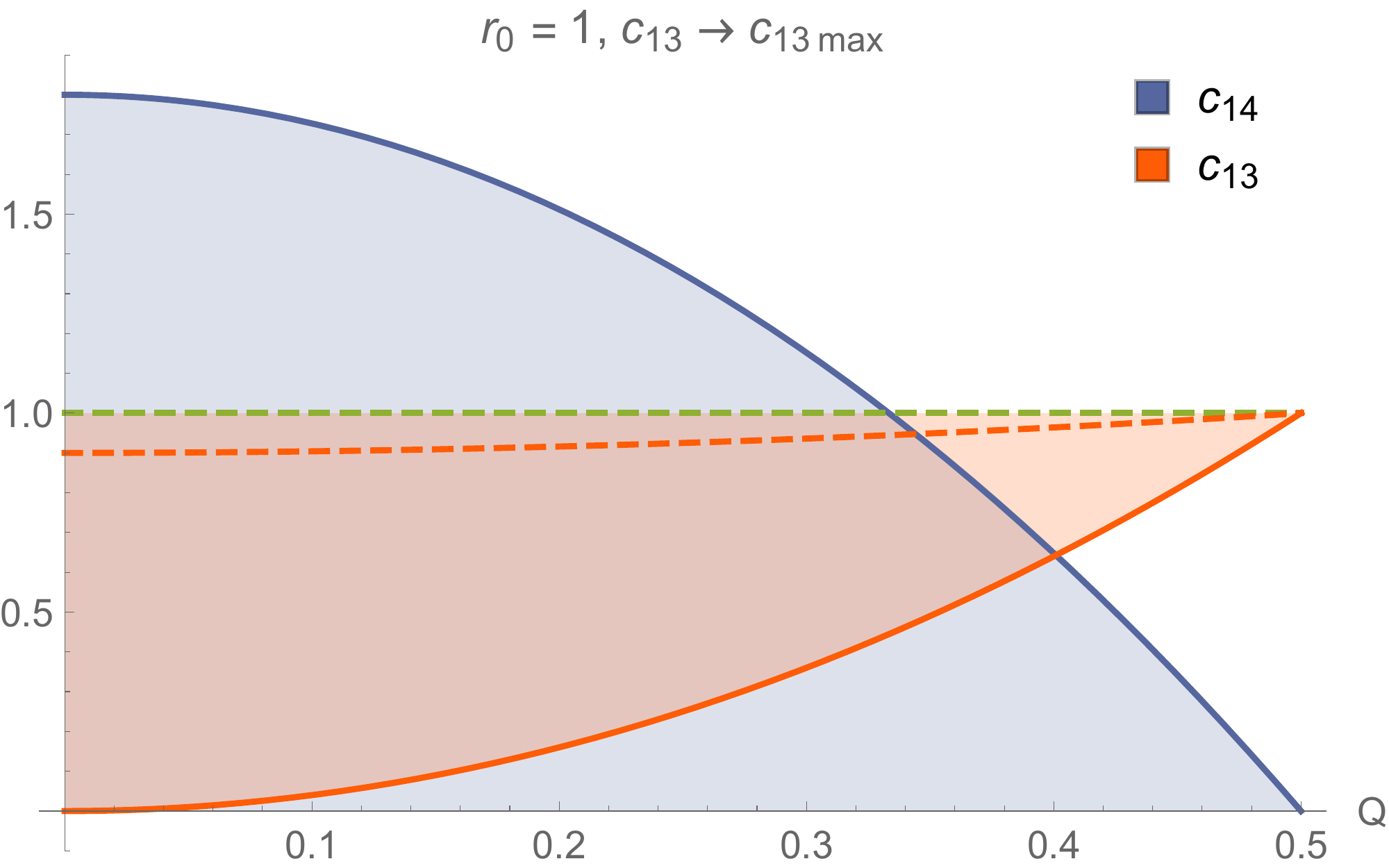}
  \includegraphics[width = 0.4\textwidth]{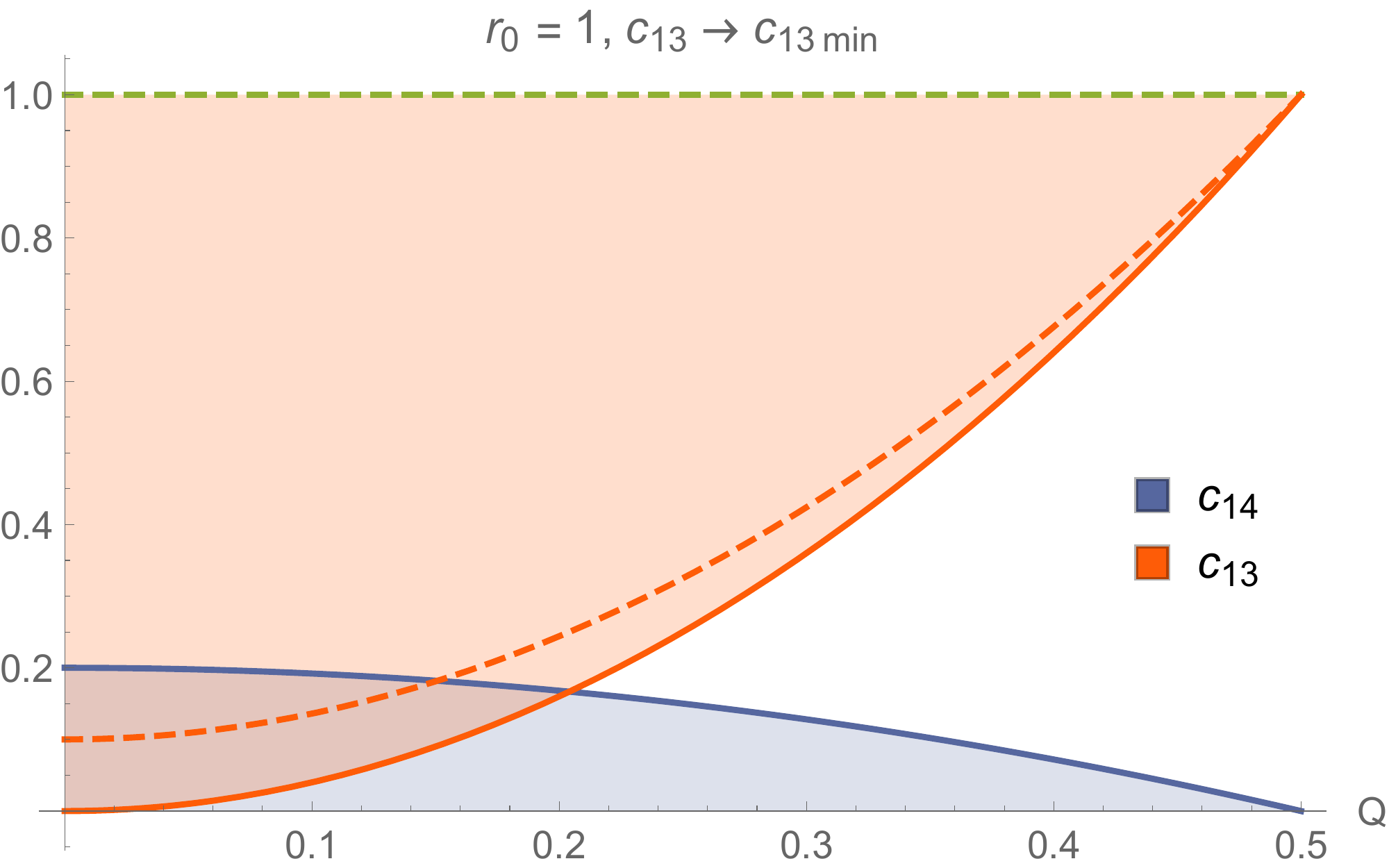}
  \caption{The allowable range of $c_{13}$ (red) and $c_{14}$ (blue) for the second kind aether black hole with different $Q$. The red dashed line denotes the actual value of $c_{13}$ which can change the size of the allowable range of $c_{14}$.}\label{fig:CS 2 region c14}
\end{figure}

\begin{figure}[htbp]
  \centering
  \includegraphics[width = 0.4\textwidth]{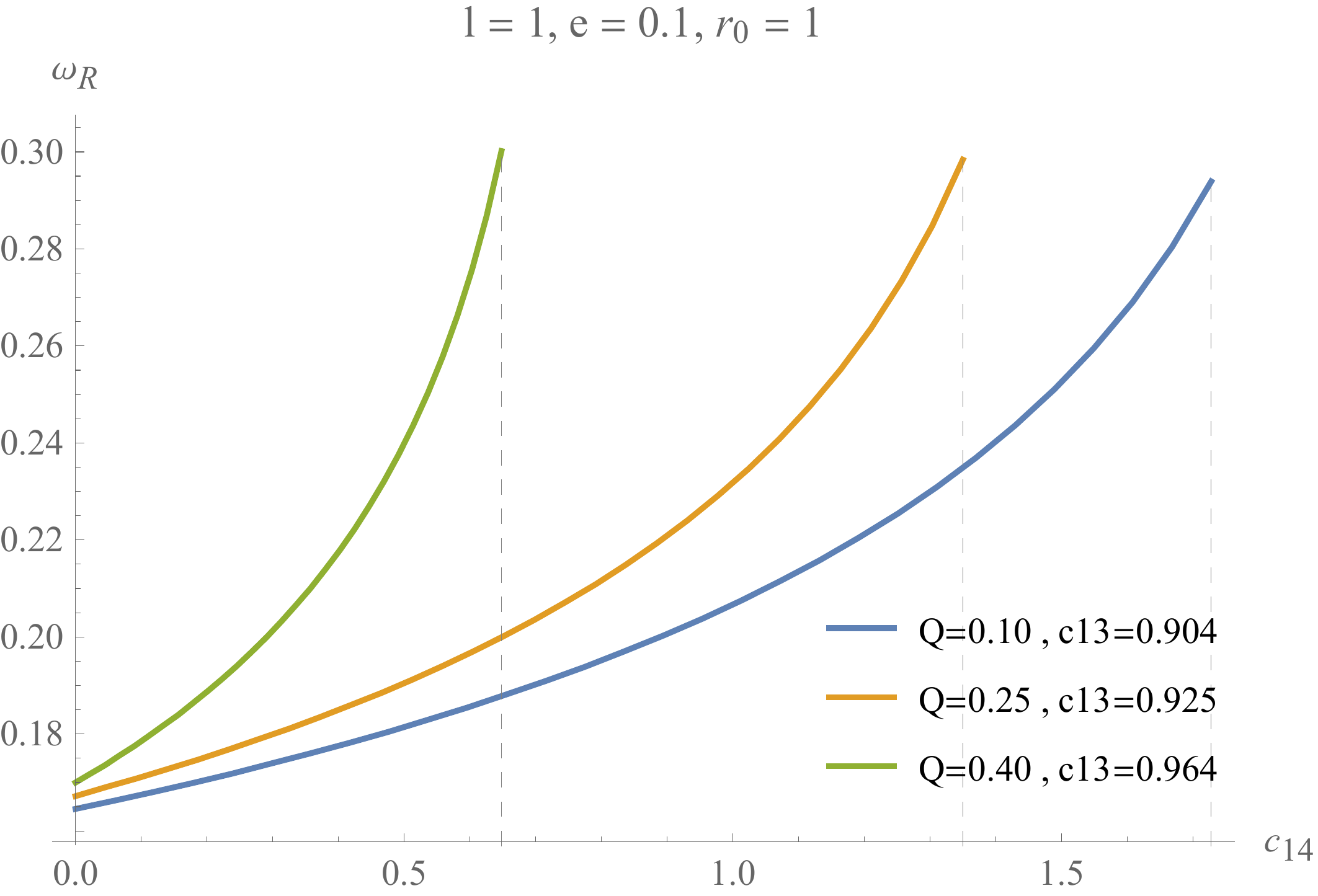}
  \includegraphics[width = 0.4\textwidth]{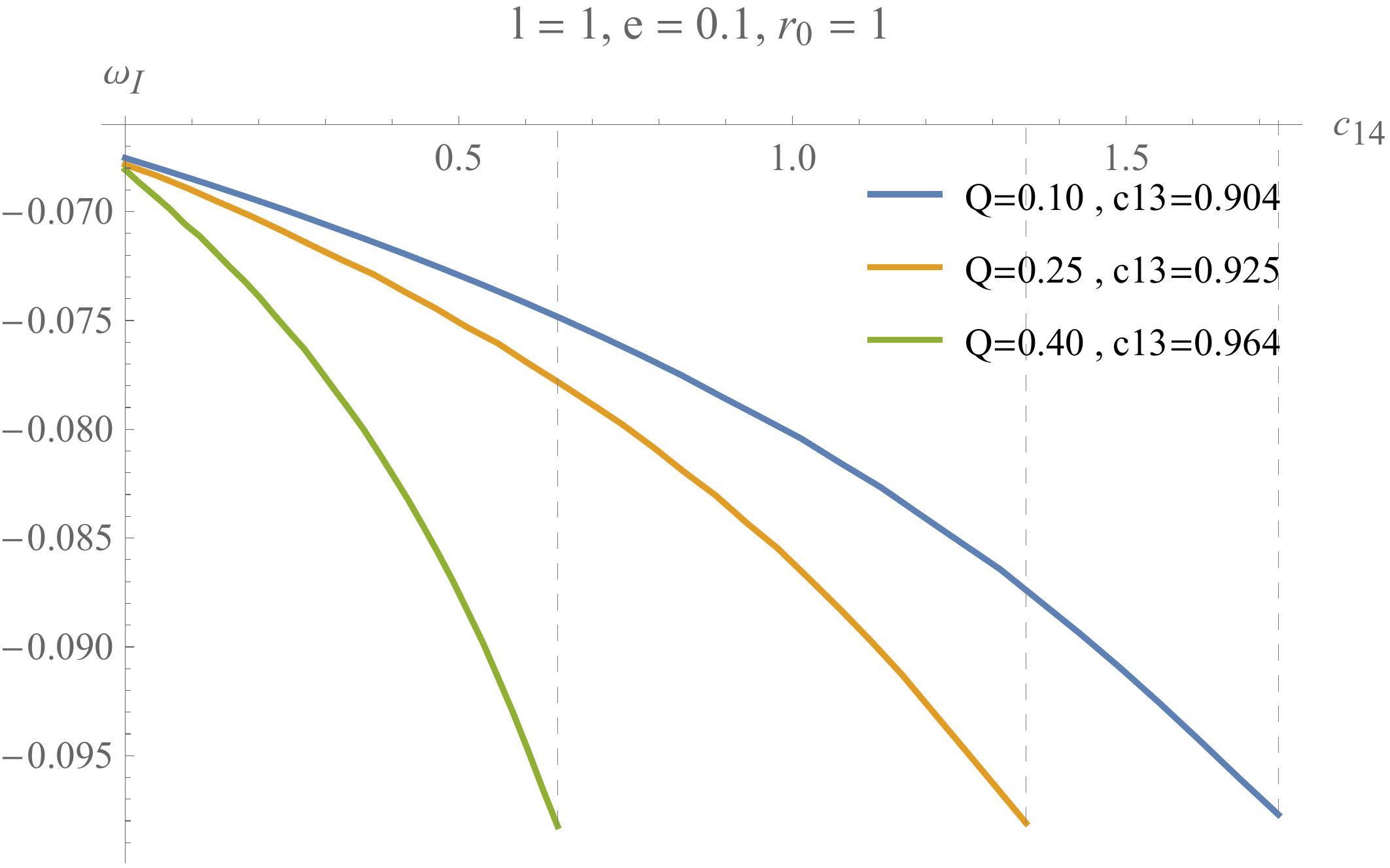}
  \caption{The left and the right plot are $\omega_{\textrm{R}}$ and $\omega_{\textrm{I}}$ of the fundamental modes vs $c_{14}$ at $l=1,\,e=0.1,\,r_{0}=2,\,c_{13}\rightarrow c_{13\:\textrm{max}}.$}
  \label{fig:CS 2 vsc14 l=1}
\end{figure}

We next show the parameters allowable range in Fig. \ref{fig:CS 2 region c14} and the fundamental modes with different $c_{14}$ in Fig. \ref{fig:CS 2 vsc14 l=1}. The parametric constraints derived from the constraints (\ref{eq:constraint for c123=0}) are given by
\begin{equation}
  \frac{4Q^{2}}{r_{0}^{2}} < c_{13} <1 ,\quad 0<c_{14}\leq\frac{-8Q^{2}+2 c_{13} r_{0}^{2}}{r_{0}^{2}}.
\end{equation}

In the same way, we choose the left plot ($c_{13} \rightarrow c_{13\: \textrm{max}}$) because of the larger allowable range of $c_{14}$. In Fig. \ref{fig:CS 2 vsc14 l=1}, the real part of frequencies increases with $c_{14}$ monotonically and the imaginary part of frequencies decreases with $c_{14}$ monotonically.

It should be noted that the modes with different fixed parameters tend to be consistent while the variable parameters tend to be the maximum value both in Fig. \ref{fig:CS 2 vsQ l=1} and \ref{fig:CS 2 vsc14 l=1}. The reason for this similar behavior is that the metric function (\ref{eq:metric function for second kind}) becomes the Schwarzschild case while the variable parameters take the maximum value.

\section{Quasi-resonance}\label{section5}
In this section we study the massive charged scalar perturbation. The general covariant equation of a massive charged scalar field is given by
\begin{equation}
  (D_{\mu}D^{\mu}-\mu^{2})\Phi=0.
  \label{eq:massive charged scalar}
\end{equation}
There is so-called quasi-resonance, which is the arbitrarily long-living mode when the field mass approaches some special values. The quasi-resonance can also be considered as the bound-state problem at the zero damping rate limit \cite{Konoplya2013}. The imaginary part of frequency $\textrm{Im}(\omega)$ increases with the increasing of the field mass $\mu$. When $\textrm{Im}(\omega)$ approaches zero, the amplitude of field function vanishes both at the horizon and infinity because of the energy conservation. Konoplya found that the existence of quasi-resonance is due to the non-zero value of potential energy at spatial infinity \cite{Konoplya2005}.

In \cite{Churilova2020}, Churilova shows that the WKB method can not be fully trusted for the calculation of quasi-resonance. Hence we choose the continued fraction method to calculate the quasi-resonance. To do this, We need to take into consideration the sub-dominant asymptotic term at infinity:
\begin{equation}
  \phi(r) \sim e^{-I(\omega)r} r^{\mu^{2}/2I(\omega)}, \quad r \rightarrow+\infty,
\end{equation}
where $\phi(r)$ is the radial part of $\Phi$ after separating variables. The following appropriate series is given by
\begin{equation}
  \phi(r) = e^{-I(\omega)\: r} \; r^{-I(\omega)+\mu^{2}/2I(\omega)} \left( \frac{r-r_{h}}{r} \right)^{H(\omega)} 
  \sum^{\infty}_{k=0} b_{k} \left( \frac{r-r_{h}}{r} \right)^{k}.
\end{equation}

\begin{figure}[htbp]
  \centering
  \includegraphics[width = 0.32\textwidth]{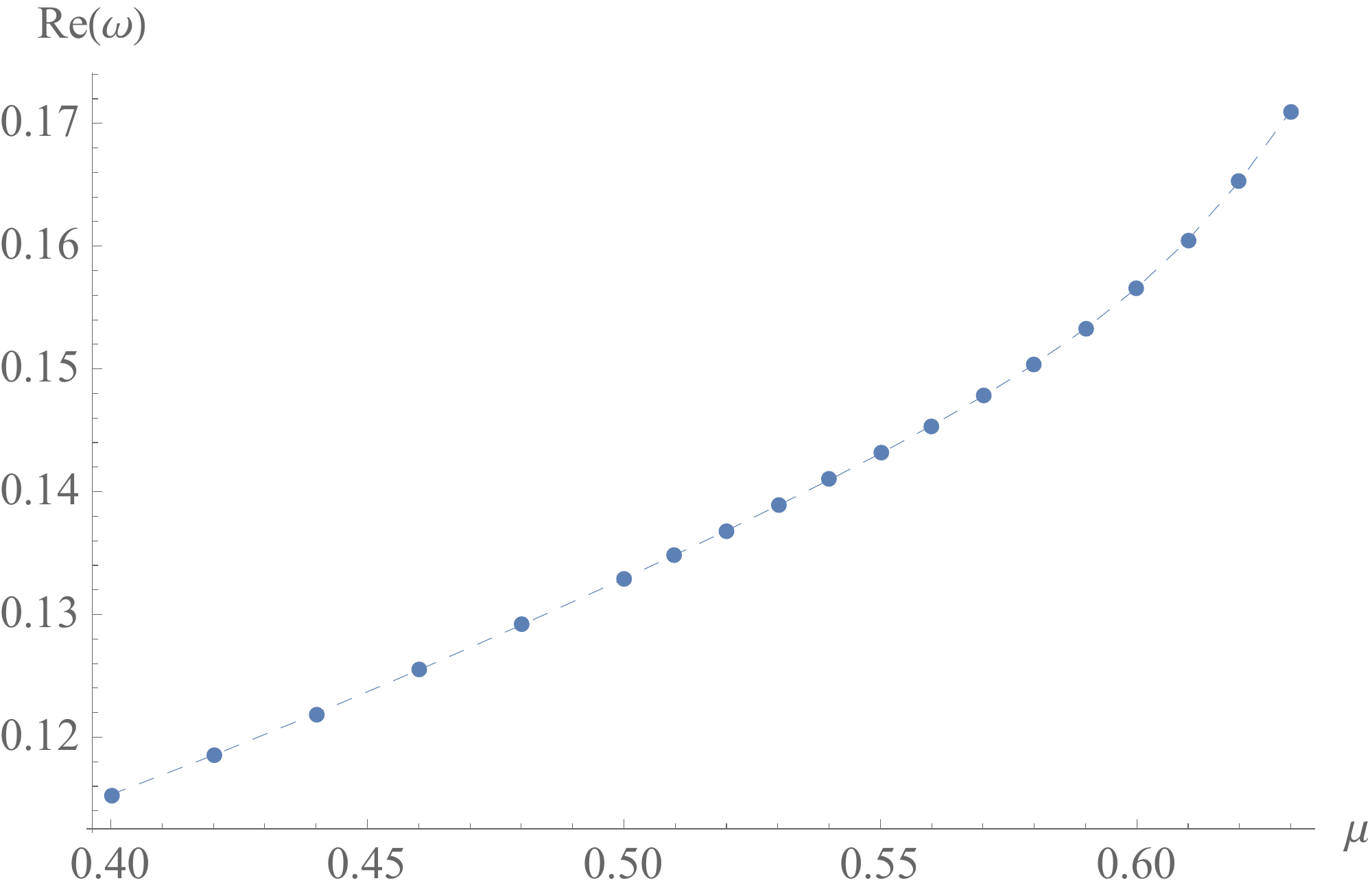}
  \includegraphics[width = 0.32\textwidth]{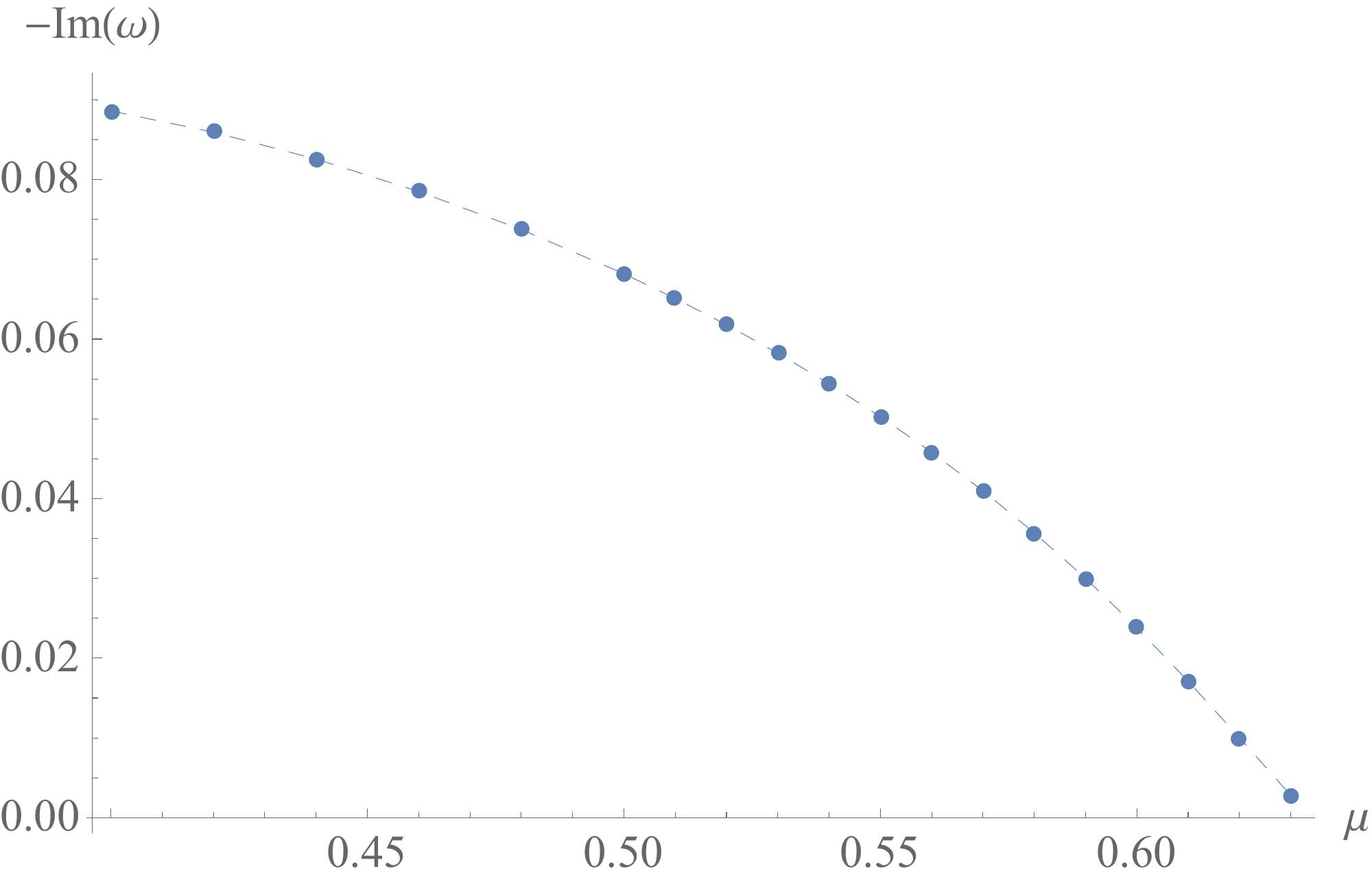}
  \includegraphics[width = 0.32\textwidth]{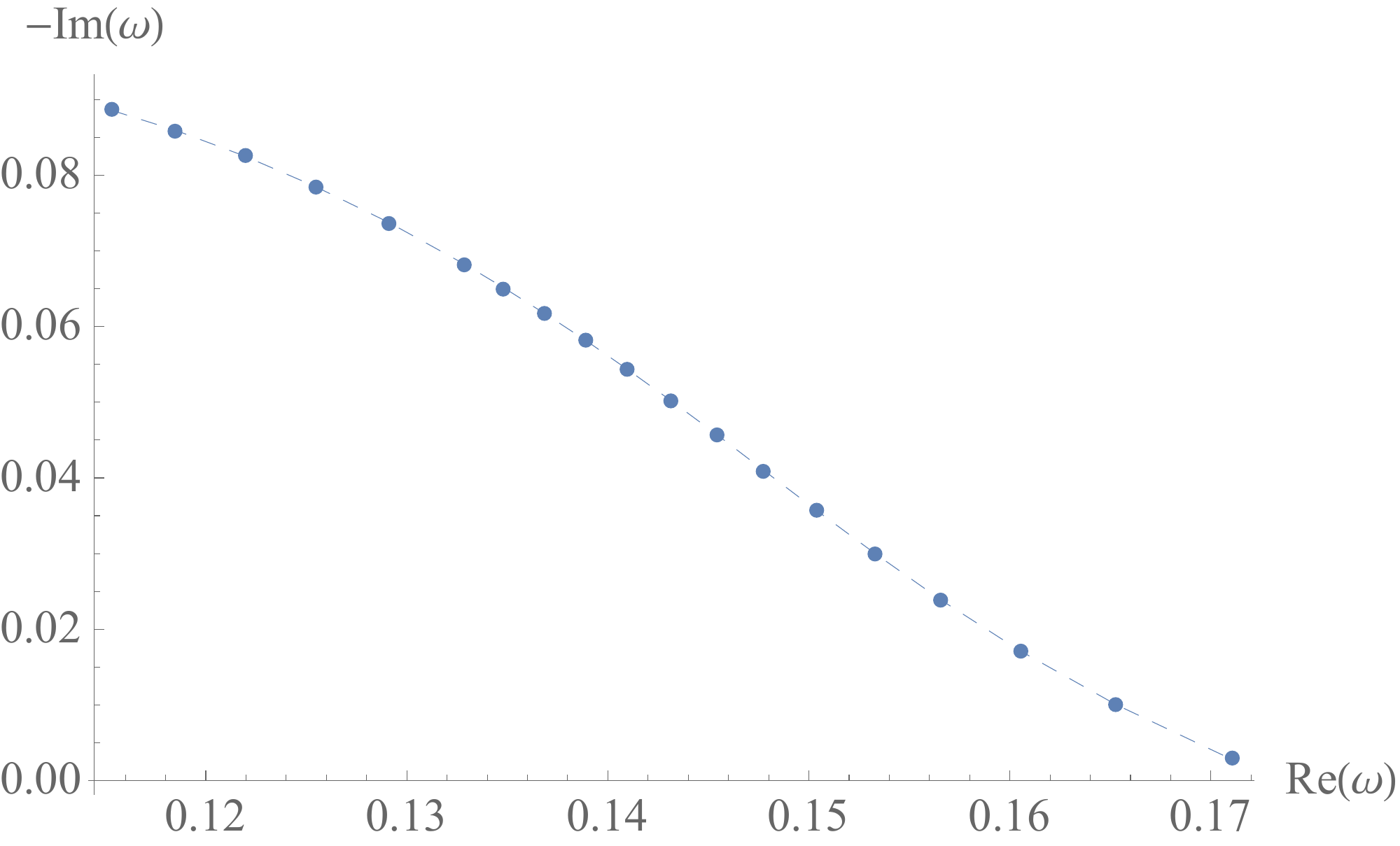}
  \caption{The fundamental modes, calculated by the continued fraction method, for the first kind aether black hole ($e=0.1,\; Q=0.1,\; r_{0}=1,\; c_{13}=0.5,\; l=0$): real parts vs $\mu$ (left panel); imaginary parts vs $\mu$ (middle panel); imaginary parts vs real parts (right panel).}\label{fig:QR c14=0 c13=0.5 l=0}
\end{figure}

\begin{figure}[htbp]
  \centering
  \includegraphics[width = 0.32\textwidth]{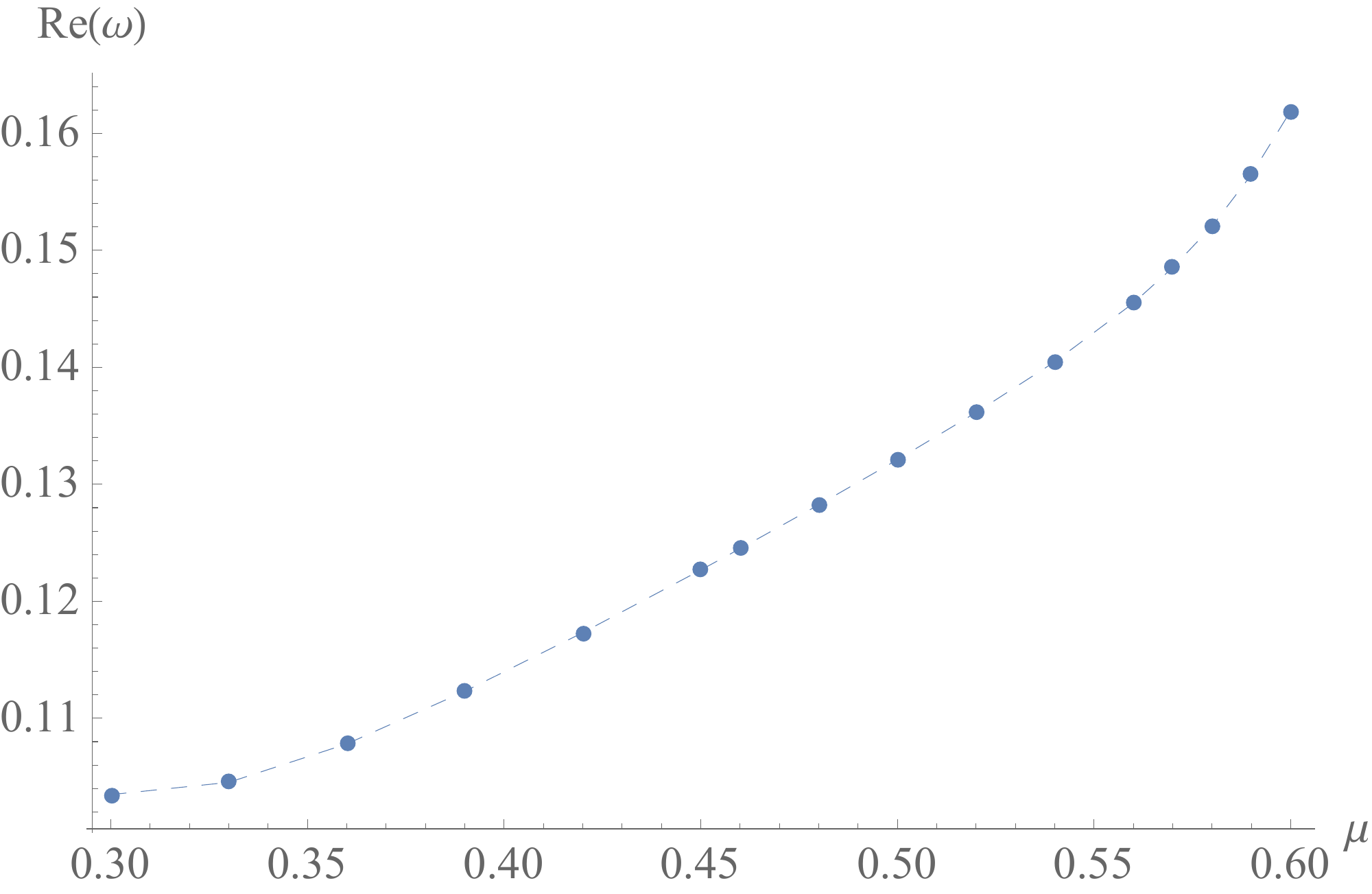}
  \includegraphics[width = 0.32\textwidth]{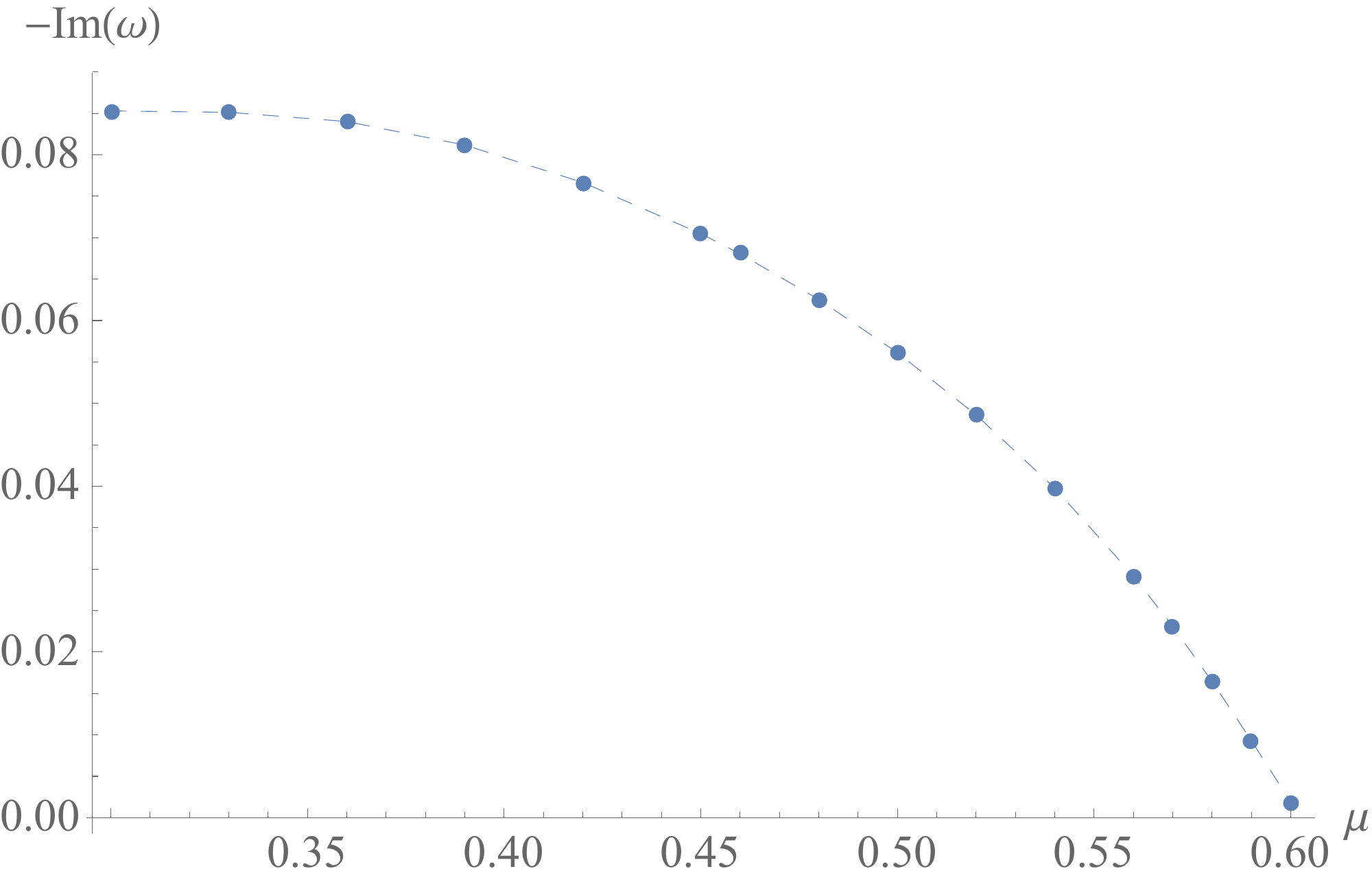}
  \includegraphics[width = 0.32\textwidth]{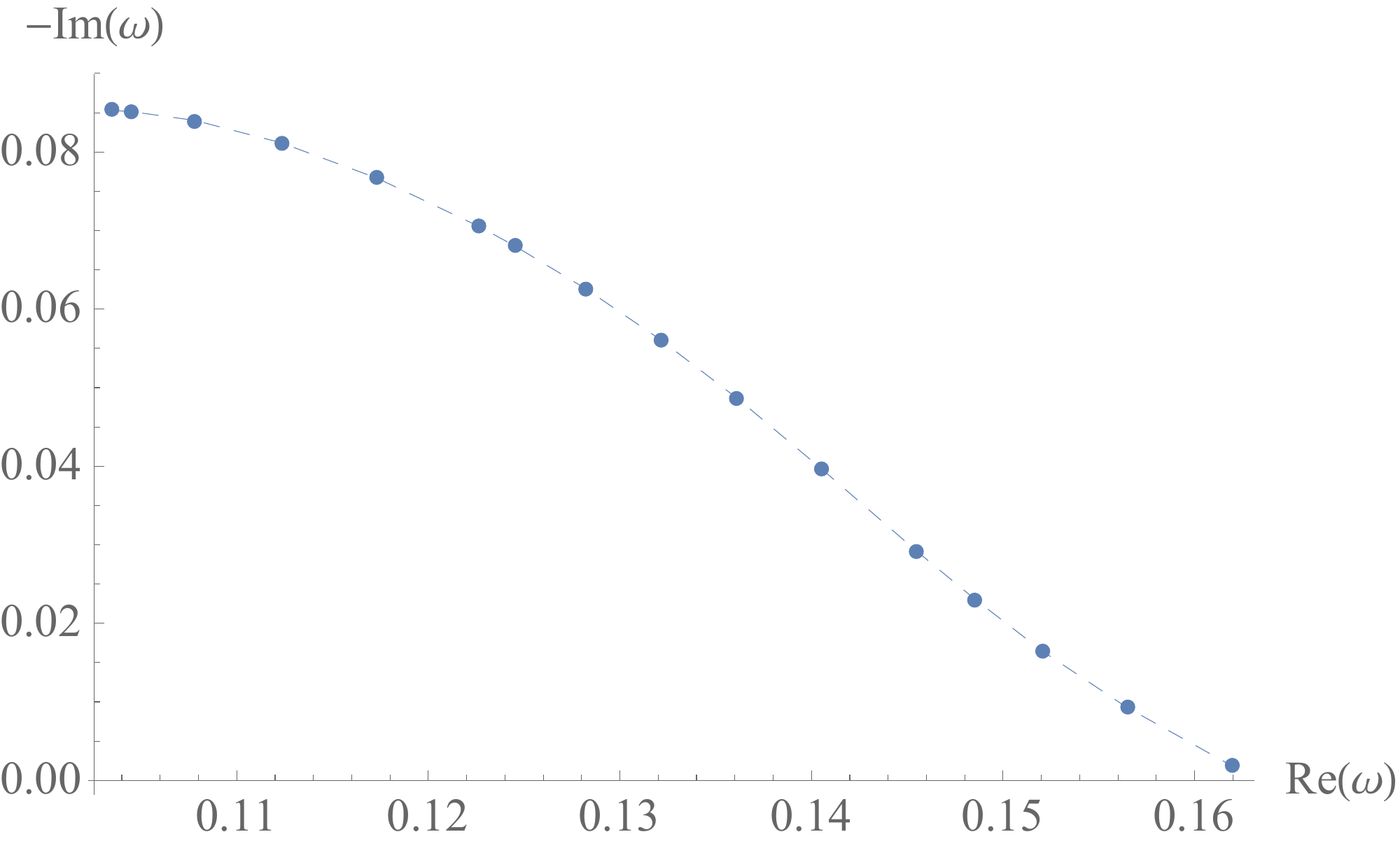}
  \caption{The fundamental modes, calculated by the continued fraction method, for the second kind aether black hole ($e=0.1,\; Q=0.1,\; r_{0}=1,\; c_{13}=0.45,\; c_{14}=0.2,\; l=0$): real part vs $\mu$ (left panel); imaginary part vs $\mu$ (middle panel); imaginary part vs real part (right panel).}\label{fig:QR c123=0 c13=0.45 c14=0.2}
\end{figure}

\begin{figure}[htbp]
  \centering
  \includegraphics[width = 0.6\textwidth]{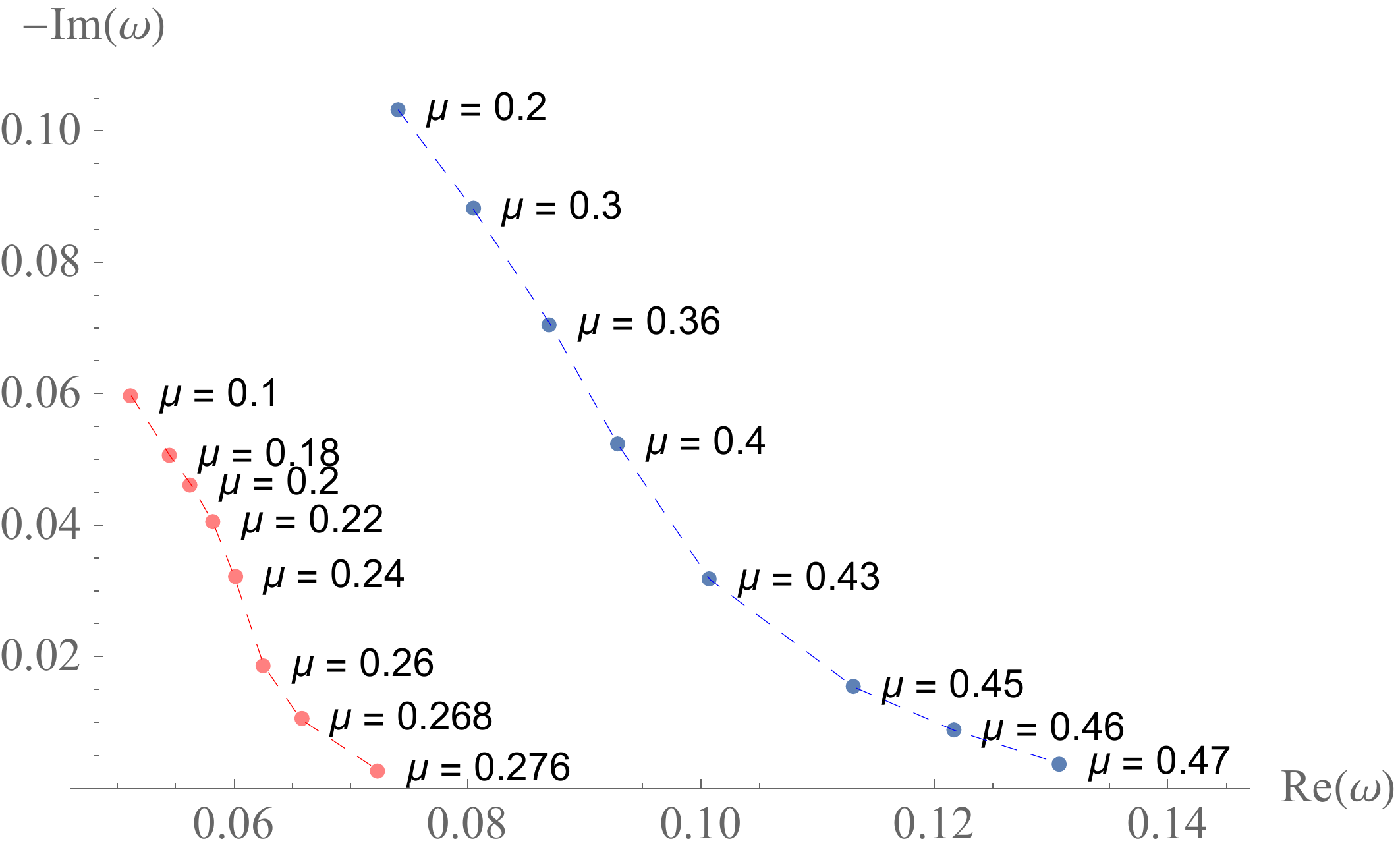}
  \caption{Dependence of the imaginary part of the fundamental mode ($e=0.1,\; Q=0.1,\; r_{0}=1,\; c_{13}=0.95,\;  l=0$) for the first kind (blue line) and the second kind (red line) aether black hole with $c_{14}=0.2$.}\label{fig:QR large c13}
\end{figure}

The Figs. \ref{fig:QR c14=0 c13=0.5 l=0}-\ref{fig:QR large c13} demonstrate that increasing of the field mass $\mu$ decreases $-\textrm{Im}(\omega)$ up to zero. We confirmed that the quasi-resonances exist for the case of the massive charged scalar field in the Einstein-Maxwell-aether black hole spacetime even with large $c_{13}$. Comparing Fig. \ref{fig:QR large c13} with Fig. \ref{fig:QR c14=0 c13=0.5 l=0} and Fig. \ref{fig:QR c123=0 c13=0.45 c14=0.2}, we find that the large $c_{13}$ decreases the critical field mass $\mu$ where $\textrm{Im}(\omega)$ approaches zero.

\section{Discussion}\label{section6}

In this paper, we studied the fundamental modes of the charged scalar perturbations in the background of two kinds of Einstein-Maxwell-aether black holes. These detailed modes with different system parameter($c_{13},Q,c_{14}$) are demonstrated by tables and figures. There are three methods, the WKB method with Pad\'{e} approximants, continued fraction method and generalized eigenvalue method used in the calculations. We extended the WKB formula with the generalized effective potential with $\omega$-dependence and proposed a new numerical program for the continued fraction method whose effectiveness has been verified in this paper. In general, the continued fraction method provides the most accurate results of quasi-normal modes. The effect of the aether parameter $c_{14}$ has not been investigated in previous studies, which is included in our content. We analyzed the allowed region obtained by the constraints of parameter for the second kind aether black hole. Finally we calculated the spectrum of the massive perturbations and confirmed the quasi-resonances in the Einstein-Maxwell-aether theory.

There are several topics worthy of further study. First we can extend the treatment of the $\omega$-dependence potential in the higher-order WKB formula, which needs to deal with the complexity and accuracy of the calculation. In addition, the gravitational perturbations of aether black hole would be more rich and interesting.

\section*{Acknowledgments}
Peng Liu would like to thank Yun-Ha Zha for her kind encouragement during this work. This work is supported by the Natural Science Foundation of China under Grant No. 11805083, 11905083, 12005077 and Guangdong Basic and Applied Basic Research Foundation (2021A1515012374).

\end{document}